\documentclass[11pt,hyper,a4paper]{article}
\usepackage{jheppub}
\usepackage{amsmath,amssymb,amsfonts,amscd,bm, amsthm, slashed, mathrsfs, bm}
\usepackage{graphicx, wrapfig}
\usepackage{multirow}
\usepackage{verbatim}
\usepackage[title]{appendix}
\usepackage{fancybox}
\usepackage[utf8]{inputenc}
\usepackage{arydshln}

\usepackage{url}
\usepackage{float}

\usepackage[normalem]{ulem}
\usepackage{graphicx}

\usepackage{color}
\usepackage[usenames,dvipsnames,svgnames,table]{xcolor}
\definecolor{darkblue}{cmyk}{0.9,0.9,0,0}
\hypersetup{colorlinks=false, linkcolor=darkblue, linkcolor=darkblue, citecolor=darkblue}

\usepackage{multirow}
\usepackage{verbatim}

\usepackage{bbm}
\usepackage{comment}
\usepackage{enumitem}
\usepackage{float}
\usepackage{empheq}
\usepackage[usenames,dvipsnames,svgnames,table]{xcolor}
\usepackage{enumerate}

\title{3d-3d Correspondence and 2d $\mathcal{N}=(0,2)$ Boundary Conditions}

\author{Hee-Joong Chung}

\affiliation{Department of Science Education, Jeju National University, Jeju, 63243, Republic of Korea}

\abstract{We consider quiver forms that appear in the motivic Donaldson-Thomas generating series or characters of conformal field theories and relate them to 3d $\mathcal{N}=2$ theories on $D^2 \times_q S^1$ with certain boundary conditions preserving 2d $\mathcal{N}=(0,2)$ supersymmetry.
We apply this to the 3d-3d correspondence and provide a Lagrangian description of 3d $\mathcal{N}=2$ theories $T[M_3]$ with 2d $\mathcal{N}=(0,2)$ boundary conditions for 3-manifolds $M_3$ in several contexts.}

\begin{document}

\maketitle

%%%%%%%%%%%%%%%%%%%%%%%%%%%%%%%%%%%%%%%%%%%%%%%%%%%%%%%%%%%%%%%%%%%%%%%%%%%%%%%%%%%%
%%%%%%%%%%%%%%%%%%%%%%%%%%%%%%%%%%%%%%%%%%%%%%%%%%%%%%%%%%%%%%%%%%%%%%%%%%%%%%%%%%%%
%%%%%%%%%%%%%%%%%%%%%%%%%%%%%%%%%%%%%%%%%%%%%%%%%%%%%%%%%%%%%%%%%%%%%%%%%%%%%%%%%%%%
%%%%%%%%%%%%%%%%%%%%%%%%%%%%%%%%%%%%%%%%%%%%%%%%%%%%%%%%%%%%%%%%%%%%%%%%%%%%%%%%%%%%
%%%%%%%%%%%%%%%%%%%%%%%%%%%%%%%%%%%%%%%%%%%%%%%%%%%%%%%%%%%%%%%%%%%%%%%%%%%%%%%%%%%%

\section{Introduction}

The 3d-3d correspondence relates complex Chern-Simons theory on a 3-manifold $M_3$ and the 3d $\mathcal{N}=2$ theory $T[M_3;G]$ that is determined by $M_3$ and the gauge group $G$ \cite{Dimofte-Gaiotto-Gukov, Dimofte-Gaiotto-Gukov-index, Chung-Dimofte-Gukov-Sulkowski}.
As Chern-Simons theory is related to various other subjects in physics and mathematics, many such subjects are also related to 3d $\mathcal{N}=2$ theories.
In particular, from recent developments in exact calculations in supersymmetric gauge theories, the 3d-3d correspondence provides new and interesting perspectives on many other new developments in physics and mathematics.
For example, the homological blocks of 3-manifolds were largely motivated by the 3d-3d correspondence \cite{Gukov-Putrov-Vafa, Gukov-Pei-Putrov-Vafa, Gukov-Manolescu}.

Among many interesting developments, a relation between the generating function of the HOMFLY polynomial and the generating function of the Donaldson-Thomas invariant or a quiver form was found \cite{KRSS1, KRSS2}, and similarly for the homological block for a knot complement \cite{Kucharski-quiver, EGGKPSS}.
It is expected that the relations hold for any knots, though they have not been proved yet.	\\

In this paper, we are interested in topological invariants of 3-manifolds that can be expressed as a quiver form or its slightly generalized versions and relate them with the 3d $\mathcal{N}=2$ theories on $D^2 \times_q S^1$ preserving 2d $\mathcal{N}=(0,2)$ supersymmetry at the boundary.
We see that such quiver forms can be expressed as the half-index of the 3d $\mathcal{N}=2$ theory with the Dirichlet boundary condition on the vector multiplets and the deformed Dirichlet boundary conditions on some chiral multiplets.
By using this relation, we obtain an explicit Lagrangian description of the 3d $\mathcal{N}=2$ theory with specific 2d $\mathcal{N}=(0,2)$ boundary conditions that corresponds to the invariant of 3-manifolds under consideration.
For example, if the conjecture that the generating function of the HOMFLY polynomial can be expressed in the form of the generating function of the Donaldson-Thomas invariant is true, then we can obtain a corresponding 3d $\mathcal{N}=2$ theory with the Dirichlet boundary conditions for a general knot.
Also, we obtain 3d $\mathcal{N}=2$ theories corresponding to the homological block for knot complements, the surgery, some closed 3-manifolds, and the colored Jones polynomials.	\\

The organization of this paper is as follows.
In section \ref{sec:q3d}, after reviewing some backgrounds, we take a generating function of the Donaldson-Thomas invariant or a quiver form and interpret it as a half-index of the 3d $\mathcal{N}=2$ theory with the Dirichlet boundary conditions preserving the 2d $\mathcal{N}=(0,2)$ supersymmetry.

In section \ref{sec:3d3dq}, we apply the dictionary obtained in section \ref{ssec:qfdbc} to topological invariants of 3-manifolds that admit the quiver form and obtain the 3d $\mathcal{N}=2$ theory whose half-index is the topological invariant of the 3-manifold under consideration.
In section \ref{ssec:lk}, we apply it to the generating function of the colored HOMFLY polynomial that arises in the large $N$ limit of the brane configuration that gives the standard 3d-3d correspondence.
Also, in section \ref{ssec:mk} we consider the homological block for a knot complement $F_K(x,a,q)$ in the large $N$ limit.
In addition, we translate the gluing formula \cite{Gukov-Manolescu} in terms of the 3d $\mathcal{N}=2$ theories when $G_{\mathbb{C}}=SL(2,\mathbb{C})$.
In section \ref{ssec:lcft}, we take two examples of closed 3-manifolds and provide 3d $\mathcal{N}=2$ theory interpretation for them where we expect that such direct translation to a 3d $\mathcal{N}=2$ theory would be possible in general.
For those examples, we also provide an interpretation on how the 3d $\mathcal{N}=2$ theory with boundary conditions change upon the reversal of the orientation of 3-manifolds or equivalently upon $q \rightarrow q^{-1}$.
We discuss that such interpretation of the 3d $\mathcal{N}=2$ theory upon the reversal of the orientation of 3-manifolds holds in a more general setup given that the 3d invariants take a quiver form and $q \rightarrow q^{-1}$ makes sense.
It is known that the colored Jones polynomial for some infinite families of knots can be expressed as a quiver form with additional $q$-Pochhammer symbols.
In section \ref{ssec:jones} we discuss the 3d $\mathcal{N}=2$ theory whose half-index is the colored Jones polynomial for some infinite families of knots.

In Appendix \ref{app:ref}, we discuss the case with the refinement and provide as an example a 3d $\mathcal{N}=2$ theory that corresponds to the $a$, $-t$-deformed homological block for the left-handed torus knot $T^{l}(2,2p+1)$ complement in $S^3$.
In Appendix \ref{app:jones}, we provide 3d $\mathcal{N}=2$ theories for the colored Jones polynomials of the twist knot $K_{-p}$ and the left-handed torus knot $T^{l}(2,2p+1)$ from the expressions of them discussed in section \ref{ssec:jones}.
In Appendix \ref{app:neumann}, we discuss an interpretation of the quiver form as the half-index of the 3d $\mathcal{N}=2$ theory with the Neumann boundary conditions on vector multiplets.
We also give some remarks on \cite{Ekholm-Kucharski-Longhi1} where partial information of the 3d $\mathcal{N}=2$ theory with the Neumann boundary conditions on vector multiplets for the quiver form was discussed.
The equivalent quivers give the same generating function of the Donaldson-Thomas invariant but with different expressions, so the 3d $\mathcal{N}=2$ theories for them are expected to be dual \cite{Ekholm-Kucharski-Longhi2, JKLNS,Cheng:2023ocj}.
In Appendix \ref{app:equivq}, we summarize the known equivalence of quivers and comment that 3d $\mathcal{N}=2$ theories whose half-indices are given by the equivalent quiver forms would be dual, which can be applied to the cases discussed in section \ref{ssec:lk} and section \ref{sssec:afk}.

%%%%%%%%%%%%%%%%%%%%%%%%%%%%%%%%%%%%%%%%%%%%%%%%%%%%%%%%%%%%%%%%%%%%%%%%%%%%%%%%%%%%
%%%%%%%%%%%%%%%%%%%%%%%%%%%%%%%%%%%%%%%%%%%%%%%%%%%%%%%%%%%%%%%%%%%%%%%%%%%%%%%%%%%%
%%%%%%%%%%%%%%%%%%%%%%%%%%%%%%%%%%%%%%%%%%%%%%%%%%%%%%%%%%%%%%%%%%%%%%%%%%%%%%%%%%%%
%%%%%%%%%%%%%%%%%%%%%%%%%%%%%%%%%%%%%%%%%%%%%%%%%%%%%%%%%%%%%%%%%%%%%%%%%%%%%%%%%%%%
%%%%%%%%%%%%%%%%%%%%%%%%%%%%%%%%%%%%%%%%%%%%%%%%%%%%%%%%%%%%%%%%%%%%%%%%%%%%%%%%%%%%

\section{Quiver forms and 3d $\mathcal{N}=2$ theories}
\label{sec:q3d}

The 3d-3d correspondence is about the relation between $G_{\mathbb{C}}$ Chern-Simons theory on a 3-manifold $M_3$ and the 3d $\mathcal{N}=2$ theory $T[M_3; G]$ which depends on topological information of $M_3$ and the gauge group $G$.
Via the correspondence, the topological quantities in the Chern-Simons theory are matched with the BPS quantities in the 3d $\mathcal{N}=2$ theory $T[M_3; G]$.
The correspondence was proposed via a bottom-up approach by using the ideal tetrahedra triangulations \cite{Dimofte-Gaiotto-Gukov, Dimofte-Gaiotto-Gukov-index}.
However, this correspondence didn't include abelian flat connections \cite{Chung-Dimofte-Gukov-Sulkowski}, which are crucial information, for example, considering recent developments on the homological block.

The homological block, which is a new topological invariant of 3-manifolds, was introduced in \cite{Gukov-Putrov-Vafa, Gukov-Pei-Putrov-Vafa, Gukov-Manolescu}.
The homological block of the 3-manifold $M_3$ is regarded as a partition function of analytically continued Chern-Simons theory and often denoted in literature by $\widehat{Z}(M_3)$, $\widehat{Z}(M_3\backslash K)$, or $F_K$ for a knot $K$.
It is the contribution from the abelian flat connections to the Chern-Simons partition function, or in other words, they are labelled by abelian flat connections.\footnote{For knot complements, there are also homological blocks or two-variable series invariants with possibly additional deformation parameters for non-abelian branches \cite{EGGKPSS}. In this paper, we focus on the homological blocks for abelian branches.} 
Interestingly, they also know the contributions from non-abelian flat connections.
This can be seen from resurgent analysis, and it was checked, for example, for an infinite family of Seifert manifolds and Seifert knots \cite{Gukov-Marino-Putrov, Andersen-Mistegard, Chung-seifert, Chung-resurg, Chung-seifertknot}.
The homological block corresponds to the $D^2 \times_q S^1$ partition function or the half-index \cite{Gadde-Gukov-Putrov-wall, Sugiyama-Yoshida, Dimofte-Gaiotto-Paquette} of a 3d $\mathcal{N}=2$ theory with a 2d $\mathcal{N}=(0,2)$ boundary condition.
Therefore, the 3d $\mathcal{N}=2$ theory whose half-index is equal to the homological block would be a theory that captures all flat connections in the 3d-3d correspondence.

The colored HOMFLY or the colored Jones polynomial of a knot $K$ can be regarded as a certain specialization of the homological block for a knot $K$.
In other words, the homological block for a knot $K$ can be regarded as the analytic continuation of the colored HOMFLY or Jones polynomial \cite{Gukov-Manolescu, Park-largecolor, Chung-resurg}.
The colored HOMFLY polynomial contains the contributions from all flat connections, and accordingly so does the generating function of it, which appears as the open topological string partition function of the resolved conifold with the Lagrangian branes.
Therefore the 3d $\mathcal{N}=2$ theory whose half-index are such quantities would also contain the contributions from all flat connections.	\\

The brane construction for the 3d-3d correspondence in M-theory is provided by
\begin{align}
\begin{tabular}{r c c c c c c}
\text{space-time}		&	&	$S^1$ 	&$\times$ 	&$TN$ 	&$\times$ 	&$T^* M_3$	\\
$N$ \text{M5 branes}		&	&	$S^1$ 	&$\times$ 	&$D^2$ 	&$\times$ 	&$M_3$	\\
$N^{\prime}$ \text{M5$^{\prime}$ branes}		&	&	$S^1$ 	&$\times$ 	&$D^2$ 	&$\times$ 	&$L_K$
\end{tabular}
\label{mconf1}
\end{align}
where $TN$ is the Taub-NUT space and $D^2$ is a disc in $TN$, $D^2 \subset TN$ \cite{Ooguri-Vafa, Witten-M5knots}.
If there is a knot $K$ in the 3-manifold $M_3$, there are additional M5 branes, say M5$^\prime$ branes, supported on $L_K$.
Here, $L_K$ is a conormal bundle of a knot $K \subset M_3$ such that $M_3 \cap L_K = K$, which is a Lagrangian submanifold in $T^* M_3$.

In this setup, the 3d-3d correspondence is about the correspondence between the 3d $\mathcal{N}=2$ theory on $D^2 \times_q S^1$ and the Chern-Simons theory on the 3-manifold $M_3$ or $M_3 \backslash K$ when there is a knot $K$.
In the presence of a knot, the correspondence from \eqref{mconf1} is often called the 3d-3d correspondence for a knot complement.

In the brane configuration \eqref{mconf1}, there are two symmetries, $U(1)_R$ and $U(1)_q$ symmetries.
$U(1)_R$ symmetry denotes the R-symmetry of the 3d $\mathcal{N}=2$ theory and $U(1)_q$ symmetry is given by a linear combination of the rotational symmetry on $D^2$ and $U(1)_R$ symmetry.
They provide two gradings of the homological invariant that gives the homological block, which is the half-index of a 3d $\mathcal{N}=2$ theory with a boundary condition $\mathcal{B}$
\begin{align}
\mathcal{I}_{\mathcal{B}}(q) = \text{Tr}_{\mathcal{H}_\mathcal{B}} (-1)^{j} q^i
\label{grhind}
\end{align}
where $\mathcal{H}_{\mathcal{B}}$ denotes the space of BPS states.
In the presence of additional M5$^{\prime}$ branes, there are at most $N^{\prime}$ additional parameters, $x_n \in \mathbb{C}^*$, $n =1, \ldots, N^{\prime}$ in \eqref{grhind}.
When $M_3$ is a Seifert manifold, there is another $U(1)_\beta$ symmetry, which arises due to the semi-free $U(1)$ action on the Seifert manifold, and this leads to an additional grading in the homological invariants (\textit{c.f.} Appendix \ref{app:ref}).	\\

It is also possible to consider another 3d-3d correspondence by taking a large $N$ limit in \eqref{mconf1}.
We take the 3-manifold $M_3$ as a three-sphere $S^3$.
With the large $N$ limit, the brane configuration \eqref{mconf1} goes through the geometric transition, and the brane configuration becomes
\begin{align}
\begin{tabular}{r c c c c c c}
\text{space-time}		&	&	$S^1$ 	&$\times$ 	&$TN$ 	&$\times$ 	&$X$	\\
$N^{\prime}$ \text{M5$^{\prime}$ branes}		&	&	$S^1$ 	&$\times$ 	&$D^2$ 	&$\times$ 	&$L_K$
\end{tabular}
\label{mconf2}
\end{align}
where $X$ is a resolved conifold $X := \mathcal{O}(-1) \oplus \mathcal{O}(-1) \rightarrow \mathbf{P}^1$.
$L_K$ denotes a knot conormal of $K$, which is a special Lagrangian submanifold in $X$ where we abuse a notation.
In this setup, one can consider the correspondence between the Chern-Simons theory on $L_K$ and the 3d $\mathcal{N}=2$ theory $T[L_K]$ on $D^2 \times_q S^1$.
It is often called the 3d-3d correspondence for a knot conormal.
There is an additional parameter $a = q^{N}$ in this configuration, which is a complexified K\"ahler parameter.
It is a fugacity for the $U(1)_a$ global symmetry that arises from the internal 2-cycle in the resolved conifold $X$.

%%%%%%%%%%%%%%%%%%%%%%%%%%%%%%%%%%%%%%%%%%%%%%%%%%%%%%%%%%%%%%%%%%%%%%%%%%%%%%%%%%%%
%%%%%%%%%%%%%%%%%%%%%%%%%%%%%%%%%%%%%%%%%%%%%%%%%%%%%%%%%%%%%%%%%%%%%%%%%%%%%%%%%%%%
%%%%%%%%%%%%%%%%%%%%%%%%%%%%%%%%%%%%%%%%%%%%%%%%%%%%%%%%%%%%%%%%%%%%%%%%%%%%%%%%%%%%

\subsection{Donaldson-Thomas invariant for symmetric quiver and quiver form}

A quiver $Q=(I,A)$ is an oriented graph where $I$ is a finite set of vertices or nodes and $A$ is a finite set of arrows $\alpha : i \rightarrow j$ for $i,j \in I$.
If $\alpha : i \rightarrow i$, then such arrow form a loop.
The number of arrows $\alpha : i \rightarrow j$ is encoded in the adjacency matrix $C_{ij}$.
For a symmetric quiver, the number of arrows from the node $i$ to the node $j$ is the same as the number of arrows from the node $j$ to the node $i$, \textit{i.e.} $C_{ij}=C_{ji}$.

A vector space $V_i$ of dimension $d_i$ is associated to each node $i \in I$, and a linear map $M_\alpha : V_i \rightarrow V_j$ is associated to the arrow $\alpha : i \rightarrow j$.
The moduli space of quiver representations is the space of such representations $((V_i)_{i \in I}, (M_\alpha:M_i \rightarrow M_j)_{\alpha : i \rightarrow j})$ with some conditions.

The motivic Donaldson-Thomas (DT) invariant captures certain topological information of the moduli space of quiver representations \cite{KS-coha}.
One can also consider a generating function for the motivic DT invariant, and it is difficult to have an explicit expression for a general quiver.
However, an explicit expression is known for symmetric quivers \cite{KS-coha, Efimov, Meinhardt-Reineke, Franzen-Reineke}.

For a symmetric quiver with $L$ nodes, the generating function of the Donaldson-Thomas invariant takes a form of
\begin{align}
\mathcal{P}_Q(\mathbf{x},q) = \sum_{\mathbf{d} \geq 0}  \frac{(-q^{\frac{1}{2}})^{ \mathbf{d} \cdot \mathbf{C} \cdot \mathbf{d} } \, \mathbf{x}^\mathbf{d}}{(q;q)_{\mathbf{d}}} 
= \sum_{d_1, \ldots d_L \geq 0} (-q^{\frac{1}{2}})^{ \sum_{i,j} C_{ij} d_i d_j}  \prod_{j=1}^{L} \frac{\mathrm{x}_j^{d_j}}{(q;q)_{d_j}}
\label{gendt}
\end{align}
where $\mathbf{d}=(d_1, \ldots, d_L)$ is the dimension vector.
$\mathbf{x}=(\mathrm{x}_1, \ldots, \mathrm{x}_L)$ are the quiver generating parameters, which can be given by the product of several other variables including $q$.
This can be expressed as
\begin{align}
\mathcal{P}_Q(\mathbf{x},q) = \prod_{\mathbf{d} \neq 0} \prod_{j \in \mathbb{Z}} \prod_{k \geq 0} (1- (-1)^{j+1} \mathbf{x}^\mathbf{d} q^{\frac{1}{2}(j+2k+1)})^{-\Omega_{\mathbf{d},j}}
\end{align}
and $\Omega_{\mathbf{d},j} := \Omega_{d_1, \ldots, d_L;j}$ are the motivic Donaldson-Thomas invariants.

From the perspective of 3d $\mathcal{N}=2$ theories, which will be discussed in section \ref{ssec:qfdbc}, we see that the generating function of the DT invariant is expressed as
\begin{align}
\mathcal{P}_Q(\mathbf{x},q) = \sum_{d_1, \ldots d_L \geq 0} q^{\frac{1}{2} \sum_{i,j} C_{ij} d_i d_j} \prod_{j=1}^{L} \frac{x_j^{d_j} (-q^{\frac{1}{2}})^{p_j d_j}}{(q;q)_{d_j}}
\label{gendtn}
\end{align}
where $p_j$, $j=1, \ldots, L$ are integers and $x_j$ doesn't depend on $q$.
In other words, it is expected that at least for the cases that are related to 3d $\mathcal{N}=2$ theories the linear dependence on $d_j$ in the exponents of $q$ and $-1$ arises in the form of $(-q^{\frac{1}{2}})^{p_j d_j}$ if $x_j$ doesn't contain $-1$.
For the cases discussed in this paper, $x_j$ takes a form of $x_j= \prod_{s} u_{s}^{e_{s,j}}$ where the exponents $e_{s,j}$ are integers.
For example, for the knot invariant that is in the form of \eqref{gendtn} whose variables are $q$ and $x$ with possibly additional deformation parameters $a$ and $-t$, $x_j$ takes a form of $x_j = x^{e_{x,j}} a^{e_{a,j}} (-t)^{e_{t,j}}$ where $e_{x,j}$, $e_{a,j}$, and $e_{t,j}$ are integers.	\\

The form of the motivic DT generating function and its variants appear in several contexts, which include knot invariants and the characters of conformal field theories.
Such form is often called the quiver form.

For a single brane supported on $L_K$ in \eqref{mconf2}, the open topological string partition function of the system \eqref{mconf2} is given by the generating function $\mathcal{P}_K(y,a,q) = \sum_{r=0}^{\infty} P^{\mathcal{S}^r}_K(a,q) y^{-r}$ of the colored HOMFLY polynomial $P^{\mathcal{S}^r}_K(a,q)$ with the totally symmetric representation $\mathcal{S}^r$.
This generating function can be understood as the Chern-Simons partition function on $L_K$ or as the $D^2 \times_q S^1$ partition function of 3d $\mathcal{N}=2$ theory $T[L_K]$ determined by $L_K$.
It was conjectured in \cite{KRSS1, KRSS2} that the HOMFLY generating function is expressed as a form of the motivic DT generating function \eqref{gendt} for a symmetric quiver.
This conjecture was proven for some infinite families of knots \cite{Stosic-Wedrich1, Stosic-Wedrich2}.
Similarly, it was conjectured in \cite{Kucharski-quiver} that the positive or negative expansion of homological block for a knot complement is also expressed as a form of \eqref{gendt}, and it was extended to non-abelian branches in \cite{EGGKPSS}.

In addition, the character of logarithmic conformal field theory (CFT) or rational CFT also takes a form that is similar to \eqref{gendt} \cite{Feigin-Feigin-Tipunin, Kedem-Klassen-McCoy-Melzer, Nahm-cft}.
In particular, from recent studies, it is shown that logarithmic CFT (log-CFT) is closely related to the homological block or the half-index of 3d $\mathcal{N}=2$ theories on $D^2 \times_q S^1$ \cite{CCFGH, CCFFGHP}.	\\

All these quantities are related to Chern-Simons partition functions or the half-index of 3d $\mathcal{N}=2$ theories, so we may consider how \eqref{gendt} or \eqref{gendtn} are realized in the context of 3d $\mathcal{N}=2$ theories.

%%%%%%%%%%%%%%%%%%%%%%%%%%%%%%%%%%%%%%%%%%%%%%%%%%%%%%%%%%%%%%%%%%%%%%%%%%%%%%%%%%%%
%%%%%%%%%%%%%%%%%%%%%%%%%%%%%%%%%%%%%%%%%%%%%%%%%%%%%%%%%%%%%%%%%%%%%%%%%%%%%%%%%%%%
%%%%%%%%%%%%%%%%%%%%%%%%%%%%%%%%%%%%%%%%%%%%%%%%%%%%%%%%%%%%%%%%%%%%%%%%%%%%%%%%%%%%

\subsection{2d $\mathcal{N}=(0,2)$ boundary conditions of 3d $\mathcal{N}=2$ theories}

We are interested in the half-index, \textit{i.e.} the partition function of 3d $\mathcal{N}=2$ theories on $D^2 \times_q S^1$ with 2d $\mathcal{N}=(0,2)$ boundary conditions \cite{Gadde-Gukov-Putrov-wall, Sugiyama-Yoshida, Dimofte-Gaiotto-Paquette, Bullimore:2020jdq}.
In this section, we briefly review the materials in \cite{Dimofte-Gaiotto-Paquette} that are relevant to later discussion.	\\

The 3d chiral multiplet $\Phi_{\text{3d}}$ contains a complex boson $\phi$, complex fermion $\psi_{\pm}$, and its conjugate $\bar{\psi}_{\pm}$ on-shell.
The Neumann (N) boundary condition on $\Phi_{\text{3d}}$ imposes
\begin{align}
\text{Neumann (N) on $\Phi_{\text{3d}}$ : }	\quad
\partial_{\perp} \phi \big|_{\partial} = 0	\,	,	\quad	\psi_{-} \big|_{\partial} = 0
\end{align}
where $\partial$ denotes the boundary and $\perp$ means the perpendicular direction to the boundary.
The Dirichlet (D) boundary condition on $\Phi_{\text{3d}}$ gives
\begin{align}
\text{Dirichlet (D) on $\Phi_{\text{3d}}$ : }	\quad
\phi \big|_{\partial} = 0	\,	,	\quad	\psi_{+} \big|_{\partial} = 0	\,	.
\end{align}
The Dirichlet boundary condition can be deformed in such a way that the complex scalar component $\phi$ takes a constant complex number $c$ at the boundary, which is called the deformed Dirichlet boundary condition and denoted by D$_c$,
\begin{align}
\text{deformed Dirichlet (D$_c$) on $\Phi_{\text{3d}}$ : }	\quad
\phi \big|_{\partial} = c	\,	,	\quad	\psi_{+} \big|_{\partial} = 0	\,	.
\label{dcbc}
\end{align}
This boundary condition \eqref{dcbc} breaks the symmetries under which the chiral multiplet $\Phi_\text{3d}$ is charged.
Thus, in order to preserve supersymmetry at the boundary while imposing the D$_c$ boundary condition on $\Phi_{\text{3d}}$, the R-charge of $\Phi_{\text{3d}}$ must be zero.
This value is under the unitary bound, so may not be appropriate for the RG flow to a superconformal boundary condition in the IR.
However, it can be allowed in the UV if the R-charge of gauge invariant operators is above the unitary bound.	\\

There is an anomaly for 2d $\mathcal{N}=(0,2)$ boundary conditions, and the gauge anomaly should be cancelled when the vector multiplet satisfies the Neumann boundary condition.
It is useful to consider anomaly polynomial for supermultiplets and the supersymmetric Chern-Simons term with given boundary conditions.	\\

The 2d $\mathcal{N}=(0,2)$ chiral and Fermi multiplet with charge $\alpha$ under $U(1)$ and $\rho$ under $U(1)_R$ contribute to the anomaly polynomial as
\begin{align}
\text{2d chiral : } \, -(\alpha \mathbf{f} + (\rho-1)\mathbf{f}_R)^2	\,	,	\qquad
\text{2d Fermi : } \, (\alpha \mathbf{f} + \rho \mathbf{f}_R)^2
\end{align}
where $\mathbf{f}$ and $\mathbf{f}_R$ denote the field strength of $U(1)$ and $U(1)_R$.
The contribution from the 2d $\mathcal{N}=(0,2)$ vector multiplet to the anomaly polynomial is $\mathbf{f}_R^2$.	\\

The 3d $U(1)$ Chern-Simons term with level $k$ contributes by $k \mathbf{f}^2$ to the anomaly polynomial.

The contributions of the 3d chiral multiplet with charge $\alpha$ under $U(1)$ and $\rho$ under $U(1)_R$ symmetry to the anomaly polynomial are given by
\begin{align}
\text{Neumann (N) : } \, -\frac{1}{2}(\alpha \mathbf{f} + (\rho -1)\mathbf{f}_R)^2	\,	,	\qquad
\text{Dirichlet (D) : } \, \frac{1}{2}(\alpha \mathbf{f} + (\rho -1)\mathbf{f}_R)^2	\,	.
\end{align}
The contribution from the 3d $U(1)$ vector multiplet is
\begin{align}
\text{Neumann ($\mathcal{N}$) : } \, \frac{1}{2} \mathbf{f}_R^2	\,	,	\qquad
\text{Dirichlet ($\mathcal{D}$) : } \, -\frac{1}{2} \mathbf{f}_R^2	\,	.
\end{align}

\vspace{5mm}

The half-index counts BPS operators that are in the cohomology of a supercharge in the 2d $\mathcal{N}=(0,2)$ superalgebra preserved by the boundary condition $\mathcal{B}$. 
It is defined as
\begin{align}
\mathcal{I}_{\mathcal{B}}(x;q) = \text{Tr}_{\mathcal{H}_\mathcal{B}} (-1)^R q^{J+ \frac{R}{2}} x^l
\label{indform}
\end{align}
where $R$ and $J$ denote the R-charge and the spin for the $U(1)$ rotation symmetry on $D^2$ in \eqref{mconf1}, respectively.\footnote{Originally, the index is given by $\text{Tr}_{\mathcal{H}_\mathcal{B}} (-1)^F q^{J+ \frac{R}{2}} x^l$ with the fermion number $F=2J$.
When the R-charges $R$ are integers, which is the case for 3d $\mathcal{N}=2$ theories discussed in this paper, the formula with $(-1)^F$ becomes the formula \eqref{indform} with $(-1)^R$ by the change of variable $q^{\frac{1}{2}} \rightarrow -q^{\frac{1}{2}}$.}
$x$ is a fugacity for $U(1)_x$ global symmetry and $l$ denotes the charge under $U(1)_x$.	\\

The contributions from the 2d $\mathcal{N}=(0,2)$ chiral and Fermi multiplet with charge $\alpha$ under $U(1)_x$ and $\rho$ under $U(1)_R$ are
\begin{align}
\text{2d chiral : } \, I^{\text{2d}, C}(x,q) = \frac{1}{\theta((-q^{\frac{1}{2}})^{\rho-1} x^{\alpha};q )}	\,	,	\qquad
\text{2d Fermi : } \, I^{\text{2d}, F}(x,q)  = \theta((-q^{\frac{1}{2}})^{\rho} x^{\alpha} ;q )
\end{align}
where 
\begin{align}
\theta(x;q) = ((-q^{\frac{1}{2}}) x;q)_\infty ((-q^{\frac{1}{2}}) x^{-1};q)_\infty	\,	.
\label{jtheta}
\end{align}

If there is a 2d $\mathcal{N}=(0,2)$ vector multiplet with $G=U(1)$, the index is given by 
\begin{align}
(q;q)_\infty^2 \oint \frac{ds}{2\pi i s} [\text{2d matter index}(s,q)]
\end{align}
where $s$ is a fugacity for $U(1)$.	\\

The contribution from the 3d chiral multiplet with charge $\alpha$ under $U(1)_x$ and $\rho$ under $U(1)_R$ symmetry to the half-index is given by
\begin{align}
\text{Neumann (N) : } \, I_{N}^{\text{3d}}(x;q) = \frac{1}{((-q^{\frac{1}{2}})^{\rho} x^{\alpha} ; q)_\infty }	\,	,	\quad
\text{Dirichlet (D) : } \, I_{D}^{\text{3d}}(x;q) = ((-q^{\frac{1}{2}})^{2-\rho} x^{-\alpha} ; q)_\infty	\,	.
\end{align}

For the Neumann boundary condition of a $U(1)$ vector multiplet, the half-index is given by
\begin{align}
(q;q)_\infty \oint \frac{ds}{2\pi i s} \, [ \text{matter index}(s;q) ]	\,	.
\label{nvint}
\end{align}

For the $U(1)$ vector multiplet that satisfies the Dirichlet boundary condition, the half-index is given by 
\begin{align}
\frac{1}{(q;q)_\infty} \sum_{m \in \mathbb{Z}} q^{\frac{1}{2}k_{\text{eff}}m^2} s^{k_{\text{eff}}m} \times [\text{matter index}(q^m s)]
\label{dvect}
\end{align}
where $m$ is the magnetic flux and $k_{\text{eff}}$ is the effective Chern-Simons level.
The gauge symmetry is broken at the boundary and becomes a global symmetry $G_{\partial}$ and the fugacity of $G_{\partial}$ is denoted by $s$.
When there is a mixed Chern-Simons coupling, for example, between $U(1)_\partial$ and $U(1)_x$ whose fugacity is $x$ with the effective Chern-Simons level $k^\text{eff}_{sx}=k^\text{eff}_{xs}$, the contribution to the half-index \eqref{dvect} is $x^{k^\text{eff}_{sx}m}$.
Similarly, the contribution to \eqref{dvect} from the mixed CS term between $U(1)_\partial$ and $U(1)_R$ with the effective level $k^{\text{eff}}_{sR}= k^{\text{eff}}_{Rs}$ is given by $(-q^{\frac{1}{2}})^{k^{\text{eff}}_{sR}m}$.
For the quiver gauge theories with $U(1)$ for all nodes, we have $e^{\frac{1}{2} \sum_{i,j} (k_{\text{eff}})_{ij} m_i m_j} \prod_{i}s_i^{(k_{\text{eff}})_{ii}m_i} \prod_{i<j} (s_i^{m_j}s_j^{m_i})^{(k_{\text{eff}})_{ij}}$ with $(k_{\text{eff}})_{ij}=(k_{\text{eff}})_{ji}$ and the factor $(q;q)_\infty^{-L}$ if the number of nodes is $L$ in \eqref{dvect}.

%%%%%%%%%%%%%%%%%%%%%%%%%%%%%%%%%%%%%%%%%%%%%%%%%%%%%%%%%%%%%%%%%%%%%%%%%%%%%%%%%%%%
%%%%%%%%%%%%%%%%%%%%%%%%%%%%%%%%%%%%%%%%%%%%%%%%%%%%%%%%%%%%%%%%%%%%%%%%%%%%%%%%%%%%
%%%%%%%%%%%%%%%%%%%%%%%%%%%%%%%%%%%%%%%%%%%%%%%%%%%%%%%%%%%%%%%%%%%%%%%%%%%%%%%%%%%%

\subsection{Quiver form and Dirichlet boundary conditions}
\label{ssec:qfdbc}

We consider the Dirichlet ($\mathcal{D}$) boundary condition for a $U(1)$ vector multiplet and the deformed Dirichlet (D$_c$) boundary condition for a 3d bulk chiral multiplet with R-charge zero and charge $-1$ under $U(1)_{\partial}$.
From \eqref{dvect}, the half-index of this theory is given by
\begin{align}
\frac{1}{(q;q)_\infty} \sum_{m \in \mathbb{Z}} q^{\frac{1}{2}k_{\text{eff}}m^2} (q^{1+m};q)_\infty (\text{other matter index}(q^m))
\label{dcchiral}
\end{align}
where due to D$_{c}$ boundary condition $U(1)_{\partial}$ is broken so $s$ is set to 1.
Since it is zero when $m \leq -1$, \eqref{dcchiral} becomes\footnote{We may also consider the case that the $U(1)_{\partial}$ charge of the chiral multiplet is $+1$. 
In this case, we have
\begin{align}
\frac{1}{(q;q)_\infty} \sum_{m \in \mathbb{Z}} q^{\frac{1}{2}k_{\text{eff}}m^2} (q^{1-m};q)_\infty (\text{other matter index}(q^m))
= \sum_{m \geq 0} q^{\frac{1}{2}k_{\text{eff}}m^2} \frac{1}{(q;q)_m} (\text{other matter index}(q^{-m})) 	\,	.
\end{align}
}
\begin{align}
\frac{1}{(q;q)_\infty} \sum_{m \geq 0} q^{\frac{1}{2}k_{\text{eff}}m^2} (q^{1+m};q)_\infty (\text{other matter index}(q^m)) = \sum_{m \geq 0} q^{\frac{1}{2}k_{\text{eff}}m^2} \frac{1}{(q;q)_m} (\text{other matter index}(q^m)) 	\,	,
\label{dcchiral1}
\end{align}
and we see that this form arises in \eqref{gendt} or \eqref{gendtn}. 	\\

If we take $x_j = x^{h_j}$ in \eqref{gendtn} for simplicity, the generating function of the DT invariant or the quiver form is expressed as
\begin{align}
\mathcal{P}_Q(\mathbf{x},q) = \sum_{d_1, \ldots d_{L} \geq 0} q^{\frac{1}{2} \sum_{i,j} C_{ij} d_i d_j} \prod_{j=1}^{L} \frac{x^{h_jd_j} (-q^{\frac{1}{2}})^{p_j d_j}}{(q;q)_{d_j}}
\label{qform}
\end{align}
Then, from the discussion above, this is the half-index of the 3d $\mathcal{N}=2$ abelian quiver gauge theory with
\begin{itemize}[leftmargin=5mm]
\item vector multiplets for $U(1)^{L}$ gauge groups with Dirichlet ($\mathcal{D}$) boundary condition
\item $L$ chiral multiplets $\Phi_{i}$, $i=1, \ldots, L$ of R-charge 0 and charge $-\delta_{ij}$ under $U(1)_j$, $j=1, \ldots, L$ of $U(1)^{L}$ with the deformed Dirichlet (D$_{c}$) boundary condition
\item (mixed) CS term between $U(1)^{L}$ and $U(1)^{L}$ with the effective level $C_{ij}$
\item mixed CS term between $U(1)^{L}$ and $U(1)_x$ with the effective level $(h_1, h_2, \ldots, h_{L})$
\item mixed CS term between $U(1)^{L}$ and $U(1)_R$ with the effective level $(p_1, p_2, \ldots, p_{L})$
\end{itemize}

These Chern-Simons couplings are the IR data except the ones between $U(1)^L$ and $U(1)_x$.
For the UV data we consider anomaly polynomials.
Contributions to the anomaly polynomial from the vector multiplets and chiral multiplets with the Dirichlet boundary conditions are 
\begin{align}
-\frac{L}{2}\mathbf{f}_R^2	\,	,	\qquad \sum_{j=1}^{L} \frac{1}{2}(-\mathbf{f}_j-\mathbf{f}_R)^2	\,	,
\end{align}
respectively, where $\mathbf{f}_j$ denotes the field strength of $U(1)_j$ at the $j$-th node of the quiver, which becomes the field strength of $U(1)_\partial$ for each $U(1)_j$ after the Dirichlet boundary condition.
So the sum of their contributions is $\sum_{j=1}^{L} \frac{1}{2}\mathbf{f}_j^2 + \mathbf{f}_j \mathbf{f}_R$.
The anomaly polynomial for the effective Chern-Simons coupling is
\begin{align}
\sum_{i,j=1}^{L} C_{ij} \mathbf{f}_i \mathbf{f}_j + 2 \mathbf{f}_x \sum_{j=1}^{L} h_j \mathbf{f}_j + 2\mathbf{f}_R \sum_{j=1}^{L} p_j \mathbf{f}_j	\,	.
\end{align}
Thus, the UV Chern-Simons coupling can be read off from
\begin{align}
\sum_{i,j=1}^{L} C_{ij} \mathbf{f}_i \mathbf{f}_j + 2 \mathbf{f}_x \sum_{j=1}^{L} h_j \mathbf{f}_j + 2\mathbf{f}_R \sum_{j=1}^{L} p_j \mathbf{f}_j - \bigg(\sum_{j=1}^{L} \frac{1}{2}\mathbf{f}_j^2 + \mathbf{f}_j \mathbf{f}_R\bigg)	\,	.
\end{align}
Therefore, the UV data of the 3d $\mathcal{N}=2$ quiver gauge theory is
\begin{itemize}[leftmargin=5mm]
\item vector multiplet for $U(1)^{L}$ gauge group with Dirichlet ($\mathcal{D}$) boundary condition
\item $L$ chiral multiplets $\Phi_{i}$, $i=1, \ldots, L$ of R-charge 0 and charge $-\delta_{ij}$ under $U(1)_j$, $j=1, \ldots, L$ of $U(1)^{L}$ with the deformed Dirichlet (D$_{c}$) boundary condition
\item (mixed) CS term between $U(1)^{L}$ and $U(1)^{L}$'s with the level $C_{ij}-\frac{1}{2}\delta_{ij}$
\item mixed CS coupling between $U(1)^{L}$ and $U(1)_x$ with the level $(h_1, \ldots, h_{L})$
\item mixed CS coupling between $U(1)^{L}$ and $U(1)_R$ with the level $(p_1-\frac{1}{2}, \ldots, p_{L}-\frac{1}{2})$
\end{itemize}

We obtained a 3d $\mathcal{N}=2$ theory whose half-index is a generating function of the DT invariants or the quiver form \eqref{qform}, and this is readily extended to the cases with more variables.	\\

In section \ref{sec:3d3dq}, we apply this dictionary to topological invariants of 3-manifolds that admit the quiver form or slightly generalized versions of it that contains additional $q$-Pochhammer symbols.
The relation between the 3d $\mathcal{N}=2$ theory and the quiver form is not limited to the cases of topological invariants of 3-manifolds and related quantities discussed in this paper, but can also be applied to other physical or mathematical systems whose quantity admits the quiver form.

%%%%%%%%%%%%%%%%%%%%%%%%%%%%%%%%%%%%%%%%%%%%%%%%%%%%%%%%%%%%%%%%%%%%%%%%%%%%%%%%%%%%
%%%%%%%%%%%%%%%%%%%%%%%%%%%%%%%%%%%%%%%%%%%%%%%%%%%%%%%%%%%%%%%%%%%%%%%%%%%%%%%%%%%%
%%%%%%%%%%%%%%%%%%%%%%%%%%%%%%%%%%%%%%%%%%%%%%%%%%%%%%%%%%%%%%%%%%%%%%%%%%%%%%%%%%%%
%%%%%%%%%%%%%%%%%%%%%%%%%%%%%%%%%%%%%%%%%%%%%%%%%%%%%%%%%%%%%%%%%%%%%%%%%%%%%%%%%%%%
%%%%%%%%%%%%%%%%%%%%%%%%%%%%%%%%%%%%%%%%%%%%%%%%%%%%%%%%%%%%%%%%%%%%%%%%%%%%%%%%%%%%

\section{3d-3d correspondence and quiver forms}
\label{sec:3d3dq}

With the dictionary in section \ref{ssec:qfdbc}, we consider the 3d-3d correspondence in several contexts, including the 3d-3d correspondence for the knot conormal, the knot complement, the surgery, the closed 3-manifold or the log-CFT, and the colored Jones polynomial.

%%%%%%%%%%%%%%%%%%%%%%%%%%%%%%%%%%%%%%%%%%%%%%%%%%%%%%%%%%%%%%%%%%%%%%%%%%%%%%%%%%%%
%%%%%%%%%%%%%%%%%%%%%%%%%%%%%%%%%%%%%%%%%%%%%%%%%%%%%%%%%%%%%%%%%%%%%%%%%%%%%%%%%%%%
%%%%%%%%%%%%%%%%%%%%%%%%%%%%%%%%%%%%%%%%%%%%%%%%%%%%%%%%%%%%%%%%%%%%%%%%%%%%%%%%%%%%

\subsection{3d $\mathcal{N}=2$ theory for knot conormal $L_K$}
\label{ssec:lk}

By using the result in the section \ref{ssec:qfdbc}, we can obtain a 3d $\mathcal{N}=2$ theory whose half-index is the HOMFLY generating function 
\begin{align}
\mathcal{P}_K(y,a,q) = \sum_{r=0}^{\infty} P^{\mathcal{S}^r}_K(a,q) y^{-r}	\,	,
\label{genhomfly}
\end{align}
which admit a quiver form \cite{KRSS1, KRSS2}. 
When there is a single M5 brane on $L_K$ in \eqref{mconf2}, \eqref{genhomfly} is a open topological string partition function or the Chern-Simons partition function on $L_K$, so the 3d $\mathcal{N}=2$ theory whose half-index is \eqref{genhomfly} is a $T[L_K]$ for the 3-manifold $L_K$.
If the conjecture that the generating function of the HOMFLY polynomial admit a quiver form in \cite{KRSS1, KRSS2} is true for every knot, then the result in section \ref{ssec:qfdbc} gives a direct way to obtain $T[L_K]$ for any knot $K$ with explicit boundary conditions.\footnote{
In \cite{Ekholm-Kucharski-Longhi1}, the 3d $\mathcal{N}=2$ theory $T[Q_K]$ for a quiver $Q_K$ with the Neumann boundary conditions on vector multiplets has been discussed, which is related to $T[L_K]$ after the specialization of parameters.
In Appendix \ref{app:neumann}, we discuss the quiver form as the half-index of the 3d $\mathcal{N}=2$ theory with the Neumann boundary conditions on vector multiplets.
}
As mentioned in section \ref{sec:q3d}, since the HOMFLY polynomial contains the information of all flat connections and the half-index of the theory $T[L_K]$ is the generating function of the HOMFLY polynomial, $T[L_K]$ captures all flat connections.	\\

Before we provide a few examples, we discuss the normalization that we choose.

%%%%%%%%%%%%%%%%%%%%%%%%%%%%%%%%%%%%%%%%%%%%%%%%%%%%%%%%%%%%%%%%%%%%%%%%%%%%%%%%%%%%
\subsubsection*{Normalization of knot polynomials}

There are several choices of the normalization of knot polynomial invariants, and different normalizations give different quiver forms.

The fully unreduced colored HOMFLY polynomial of the unknot for the totally symmetric representation $\mathcal{S}^r$ is given by \cite{FGS-VC}
\begin{align}
\text{fully unreduced normalization :} \qquad \overline{P}_{0_1}^{\mathcal{S}^r}(a,q) =  a^{-\frac{r}{2}} q^{\frac{r}{2}} \frac{(a;q)_r}{(q;q)_r}	\,	.
\label{unr01homfly}
\end{align}
This gives the colored Jones polynomial of the unknot $\frac{q^{\frac{r+1}{2}} - q^{-\frac{r+1}{2}} }{q^{\frac{1}{2}} - q^{-\frac{1}{2}}}$ when $a=q^2$.

Another choice of normalizations is the reduced version of knot polynomials.
In this normalization, the HOMFLY polynomial of the unknot is given by
\begin{align}
\text{reduced normalization :} \qquad P_{0_1}^{\mathcal{S}^r}(a,q) =  1	\,	.
\label{r01homfly}
\end{align}
In this section, we choose a version of unreduced HOMFLY polynomial $\widetilde{P}^{\mathcal{S}^r}(a,q)$ of the unknot, 
\begin{align}
\widetilde{P}_{0_1}^{\mathcal{S}^r}(a,q) =  \frac{(a;q)_r}{(q;q)_r}	\,	,
\label{vunr01homfly}
\end{align}
which was also used in \cite{Ekholm-Kucharski-Longhi1}, where we omit the factor $a^{-\frac{r}{2}} q^{\frac{r}{2}}$ in the fully unreduced normalization \eqref{unr01homfly}.

%%%%%%%%%%%%%%%%%%%%%%%%%%%%%%%%%%%%%%%%%%%%%%%%%%%%%%%%%%%%%%%%%%%%%%%%%%%%%%%%%%%%
\subsubsection*{Unknot $0_1$}

The generating function $\widetilde{\mathcal{P}}_{0_1}(y,a,q)$ of the colored HOMFLY polynomial $\widetilde{P}_{0_1}^{\mathcal{S}^r}(a,q)$ of the unknot is given by
\begin{align}
\widetilde{\mathcal{P}}_{0_1}(y,a,q) = \sum_{r=0}^{\infty} \frac{(a;q)_r}{(q;q)_r} y^{-r}	\,	.
\label{punr01}
\end{align}

By using identities \cite{KRSS2}
\begin{align}
\frac{(x;q)_{d_1 + \ldots + d_l}}{(q;q)_{d_1} \ldots (q;q)_{d_n}} = \sum_{\alpha_1 + \beta_1 =d_1} \cdots \sum_{\alpha_n + \beta_n =d_n} \frac{x^{\sum_{j=1}^{n} \alpha_j} q^{\frac{1}{2}\sum_{j=1}^{n} \alpha_j^2} \, q^{\sum_{l=1}^{n-1} \alpha_{l+1} \sum_{j=1}^{l} d_j} \, (-q^{\frac{1}{2}})^{-\sum_{j=1}^{n} \alpha_j}     }{\prod_{j=1}^{n}(q;q)_{\alpha_j} (q;q)_{\beta_j}}	\,	,
\end{align}
\eqref{punr01} can be expressed \cite{Ekholm-Kucharski-Longhi1} as
\begin{align}
\widetilde{\mathcal{P}}_{0_1}(y,a,q) = \sum_{d_1, d_2 \geq 0} q^{\frac{1}{2}d_1^2} 
\Bigg( \prod_{j=1}^{2} \frac{1}{(q;q)_{d_j}} \Bigg)
y^{-(d_1+d_2)} a^{d_1} (-q^{\frac{1}{2}})^{-d_1} 	\,	.
\label{homflygen01}
\end{align}
If including the prefactor $a^{-\frac{r}{2}} q^{\frac{r}{2}}$, we have an additional factor 
\begin{align}
a^{-\frac{d_1+d_2}{2}} q^{\frac{d_1+d_2}{2}}
\end{align}
in \eqref{homflygen01}, but we don't include them in \eqref{homflygen01}.\footnote{In order for \eqref{homflygen01} to fit in the form \eqref{gendtn}, additional factor $(-1)^r=(-1)^{d_1+d_2}$ is needed, which leads to $a^{-\frac{d_1+d_2}{2}} (-q^{\frac{1}{2}})^{d_1+d_2}$.}	\\

The 3d $\mathcal{N}=2$ theory $T[L_{K}]$ for the unknot can be read off from \eqref{homflygen01}, which is
\begin{itemize}[leftmargin=5mm]
\item $U(1)^2$ vector multiplets with the $\mathcal{D}$ boundary condition
\item 2 chiral multiplets $\Phi_i$, $i=1,2$ of R-charge 0 and charge $-\delta_{ij}$ under $U(1)_j$, $j=1,2$ with the D$_{c}$ boundary condition
\item UV (mixed) CS levels for $U(1)^2$ $-$ $U(1)^2$ : $
\begin{pmatrix}
\frac{1}{2}	&0	\\
0		&-\frac{1}{2}
\end{pmatrix}
$
\item UV mixed CS levels for $U(1)^2$ $-$ $U(1)_y$, $U(1)_a$, and $U(1)_R$ :
\begin{align*}
\left(-1,-1\right)	\,	,	\qquad	\left(0,1\right)	\,	,	\qquad	\left(-\frac{3}{2},-\frac{1}{2}\right)	\,	.
\end{align*}
\end{itemize}

%%%%%%%%%%%%%%%%%%%%%%%%%%%%%%%%%%%%%%%%%%%%%%%%%%%%%%%%%%%%%%%%%%%%%%%%%%%%%%%%%%%%
\subsubsection*{Trefoil knot $3^{l}_1$ (left-handed)}

The reduced colored HOMFLY polynomial for trefoil knot is given by \cite{FGS-superA}
\begin{align}
P_{3_1}^{\mathcal{S}^r}(a,q) = a^{2r} q^{-2r} \sum_{k=0}^r \frac{(q;q)_r}{(q;q)_{k} (q;q)_{r-k}} q^{2k(r+1)} \sum_{j=1}^{k} (1-a^2 q^{2(j-2)})	\,	.
\label{h31r}
\end{align}
Therefore, in the normalization \eqref{unr01homfly}, the generating function of \eqref{h31r} is \cite{Ekholm-Kucharski-Longhi1}
\begin{align}
\begin{split}
\widetilde{P}_{3_1}(y,a,q) = \sum_{\mathbf{d} \geq 0}  q^{\frac{1}{2} \sum_{i,j=1} C_{ij} d_i d_j} 
\Bigg(\prod_{j=1}^6 \frac{1}{(q;q)_{d_j}} \Bigg)
y^{-\sum_{j=1}^6 d_j}
a^{d_1+2d_2+d_3+2d_4+2d_5+3d_6}
(-q^{\frac{1}{2}})^{2d_1-3d_2-2d_4-3d_5-4d_6}
\label{homflygen31}
\end{split}
\end{align}
where
\begin{equation}
C = 
\begin{pmatrix} 
0	&0	&1	&2	&1	&2	\\
0	&1	&1	&2	&1	&2	\\
1	&1	&2	&2	&2	&3	\\
2	&2	&2	&3	&2	&3	\\
1	&1	&2	&2	&3	&3	\\
2	&2	&3	&3	&3	&4	
\end{pmatrix}	\,	.
\end{equation}

Thus the 3d $\mathcal{N}=2$ theory $T[L_K]$ whose half-index is \eqref{homflygen31} is given by
\begin{itemize}[leftmargin=5mm]
\item $U(1)^6$ vector multiplets with $\mathcal{D}$ boundary condition
\item 6 chiral multiplets $\Phi_i$, $i=1, \ldots, 6$, of R-charge 0 and charge $-\delta_{ij}$ under $U(1)_j$, $j=1, \ldots, 6$ with the D$_{c}$ boundary condition 
\item UV (mixed) CS levels for $U(1)^6$ $-$ $U(1)^6$ : $
\begin{pmatrix} 
-\frac{1}{2}	&0	&1	&2	&1	&2	\\
0	&\frac{1}{2}	&1	&2	&1	&2	\\
1	&1	&\frac{3}{2}	&2	&2	&3	\\
2	&2	&2	&\frac{5}{2}	&2	&3	\\
1	&1	&2	&2	&\frac{5}{2}	&3	\\
2	&2	&3	&3	&3	&\frac{7}{2}	
\end{pmatrix}
$
\item UV mixed CS levels for $U(1)^6$ $-$ $U(1)_y$, $U(1)_a$, and $U(1)_R$ :
\begin{align*}
\left(-1,-1,-1,-1,-1,-1\right)	\,	,	\quad	
\left(1,2,1,2,2,3\right)	\,	,	\quad	
\left(\frac{3}{2},-\frac{7}{2},-\frac{1}{2},-\frac{5}{2},-\frac{7}{2},-\frac{9}{2}\right)	\,	.
\end{align*}
\end{itemize}

%%%%%%%%%%%%%%%%%%%%%%%%%%%%%%%%%%%%%%%%%%%%%%%%%%%%%%%%%%%%%%%%%%%%%%%%%%%%%%%%%%%%
\subsubsection*{Figure-eight knot $4_1$}

The reduced colored HOMFLY polynomial of the figure-eight knot is \cite{FGS-superA} 
\begin{align}
P_{4_1}^{\mathcal{S}^r} (a,q) = \sum_{k=0}^{r} = \frac{(q;q)_r}{(q;q)_{k} (q;q)_{r-k}} a^{k} q^{k^2-k} (a^{-1} q;q^{-1})_k (a^{-1} q^{-r}; q^{-1})_k	\,	.
\end{align}
So, the generating function is 
\begin{align}
\begin{split}
&\widetilde{P}_{4_1}(y,a,q) = \sum_{\mathbf{d} \geq 0} q^{\frac{1}{2} \sum_{i,j=1}^{10} C_{ij}d_i d_j}  
\Bigg( \prod_{j=1}^{10} \frac{1}{(q;q)_{d_j}} \Bigg)	\\
&\hspace{25mm}\times y^{-\sum_{j=1}^{10} d_j}	
a^{-d_1+ d_4+ d_6 + d_8 + d_9 + 2 d_{10}}
(-q^{\frac{1}{2}})^{2m_1 - m_2 +m_3 - 2m_4 - m_6 + m_7 - 2m_8 -3 m_{10}}
\end{split}\label{homflygen41}
\end{align}
where
\begin{equation}
C=
\begin{pmatrix}
 -2 & -2 & -2 & -2 & -1 & -1 & -1 & -1 & -1 & -1 \\
 -2 & -1 & -1 & -1 & -1 & -1 & 0 & 0 & 1 & 0 \\
 -2 & -1 & -1 & -1 & 0 & 0 & 0 & 0 & 0 & 0 \\
 -2 & -1 & -1 & 0 & 0 & 0 & 0 & 1 & 1 & 1 \\
 -1 & -1 & 0 & 0 & 0 & 0 & 0 & 0 & 1 & 1 \\
 -1 & -1 & 0 & 0 & 0 & 1 & 0 & 0 & 1 & 1 \\
 -1 & 0 & 0 & 0 & 0 & 0 & 1 & 1 & 1 & 1 \\
 -1 & 0 & 0 & 1 & 0 & 0 & 1 & 2 & 2 & 2 \\
 -1 & 1 & 0 & 1 & 1 & 1 & 1 & 2 & 2 & 2 \\
 -1 & 0 & 0 & 1 & 1 & 1 & 1 & 2 & 2 & 3 
 \end{pmatrix}	\,	.
\end{equation}

Therefore, the 3d $\mathcal{N}=2$ theory $T[L_K]$ for the figure-eight knot from \eqref{homflygen41} is
\begin{itemize}[leftmargin=5mm]
\item $U(1)^{10}$ vector multiplets with the $\mathcal{D}$ boundary condition
\item 10 chiral multiplets $\Phi_i$, $i=1, \ldots, 10$ of R-charge 0 and charged $-\delta_{ij}$ under $U(1)_j$, $j=1, \ldots, 10$ with the D$_{c}$ boundary condition
\item UV (mixed) CS levels for $U(1)^{10}$ $-$ $U(1)^{10}$ : $
\begin{pmatrix} 
 -\frac{5}{2} & -2 & -2 & -2 & -1 & -1 & -1 & -1 & -1 & -1 \\
 -2 & -\frac{3}{2} & -1 & -1 & -1 & -1 & 0 & 0 & 1 & 0 \\
 -2 & -1 & -\frac{3}{2} & -1 & 0 & 0 & 0 & 0 & 0 & 0 \\
 -2 & -1 & -1 & -\frac{1}{2} & 0 & 0 & 0 & 1 & 1 & 1 \\
 -1 & -1 & 0 & 0 & -\frac{1}{2} & 0 & 0 & 0 & 1 & 1 \\
 -1 & -1 & 0 & 0 & 0 & \frac{1}{2} & 0 & 0 & 1 & 1 \\
 -1 & 0 & 0 & 0 & 0 & 0 & \frac{1}{2} & 1 & 1 & 1 \\
 -1 & 0 & 0 & 1 & 0 & 0 & 1 & \frac{3}{2} & 2 & 2 \\
 -1 & 1 & 0 & 1 & 1 & 1 & 1 & 2 & \frac{3}{2} & 2 \\
 -1 & 0 & 0 & 1 & 1 & 1 & 1 & 2 & 2 & \frac{5}{2} \\
\end{pmatrix}
$
\item UV mixed CS levels for $U(1)^{10}$ $-$ $U(1)_y$, $U(1)_a$, and $U(1)_R$
\begin{align*}
\big(\underbrace{-1, \ldots -1}_{10}\big)	\,	,	\
\left(-1,0,0,1,0,1,0,1,1,2\right)	\,	,	\
\left(\frac{3}{2},-\frac{3}{2},\frac{1}{2},-\frac{5}{2},-\frac{1}{2},-\frac{3}{2},\frac{1}{2},-\frac{5}{2},-\frac{1}{2},-\frac{7}{2}\right)	\,	.
\end{align*}
\end{itemize}

%%%%%%%%%%%%%%%%%%%%%%%%%%%%%%%%%%%%%%%%%%%%%%%%%%%%%%%%%%%%%%%%%%%%%%%%%%%%%%%%%%%%
%%%%%%%%%%%%%%%%%%%%%%%%%%%%%%%%%%%%%%%%%%%%%%%%%%%%%%%%%%%%%%%%%%%%%%%%%%%%%%%%%%%%
%%%%%%%%%%%%%%%%%%%%%%%%%%%%%%%%%%%%%%%%%%%%%%%%%%%%%%%%%%%%%%%%%%%%%%%%%%%%%%%%%%%%

\subsection{3d $\mathcal{N}=2$ theory for knot complement and surgery}
\label{ssec:mk}

In this section, we consider the homological block for a knot complement $S^3 \backslash K$.
The homological block $F_K^{SL(N,\mathbb{C})}(x,q)$ with a gauge group $G_{\mathbb{C}}=SL(N,\mathbb{C})$ is a two-variable series invariant for a knot $K$ \cite{Gukov-Manolescu, Park-largecolor}. 
When $x=q^r$, $r \in \mathbb{Z}_{\geq0}$, $F_K^{SL(N,\mathbb{C})}(x,q)$ gives an $SU(N)$ version of the colored Jones polynomial or the HOMFLY polynomial at $a=q^N$ with the totally symmetric representation $\mathcal{S}^r$.
The $a$-deformed homological block $F_K(x,a,q)$ for a knot complement $S^3 \backslash K$ was discussed in \cite{EGGKPS, EGGKPSS} and upon $a=q^N$ $F_K^{SL(N,\mathbb{C})}(x,q)$ is recovered.	\\

As in the case of the HOMFLY polynomial, we consider the normalization for the homological block $F_K$ for a knot complement since the quiver form depends on the normalization.

The fully unreduced normalization is given by
\begin{align}
\text{fully unreduced normalization :} \qquad \overline{F}_{0_1}(x,a,q) = e^{- \frac{\log x \, \log a}{2 \hbar}} x^{\frac{1}{2}} \frac{(a;q)_\infty (xq;q)_\infty}{(xa;q)_\infty (q;q)_\infty}	\,	.
\end{align}
When $a=q^2$, $x=q^r$, this gives a colored Jones polynomial of the unknot $\frac{q^{\frac{r+1}{2}} - q^{-\frac{r+1}{2}}}{q^{\frac{1}{2}} - q^{-\frac{1}{2}}}$.
Also, the reduced normalization is given by
\begin{align}
\text{reduced normalization :} \qquad F_{0_1}(x,a,q) = 1	\,	.
\label{rfk01}
\end{align}
We will use the reduced normalization \eqref{rfk01} for examples below except the unknot.

For the unknot, we choose a version of unreduced normalization \cite{EGGKPS,EGGKPSS}
\begin{align}
F_{0_1}^{u.r.}(x,a,q) = \frac{(xq;q)_\infty}{(xa;q)_\infty}	\,	.
\label{unrfk01}
\end{align}
When $a=q^2$, this gives $1-xq$.	\\

We also note that two types of expansions with respect to the variable $x$ are available for the homological block for a knot complement.
The homological block for a knot complement $S^3 \backslash K$ with the reduced normalization in \cite{Gukov-Manolescu, Park-largecolor} is a Weyl invariant homological block, which is called the balanced expansion.
They are analytically continued Chern-Simons partition functions.
However, for example, when $G_{\mathbb{C}}=SL(2,\mathbb{C})$ one can also consider a positive expansion $F_K(x,q)$ of the homological block around $x=0$ as in \cite{EGGKPS,EGGKPSS}.
By using the Weyl symmetry, one can get the negative expansion $F_K(q^{-2}x^{-1},q)$, which is obtained by the expansion around $x=\infty$, from the positive expansion $F_K(x,q)$.
By averaging them, it is possible to obtain the balanced expansion $\mathcal{F}_K(x,q)$ of a homological block.

In addition, it should be noted that in order to obtain the correct balanced expansion $\mathcal{F}_K(x,q)$ from the average of the positive and negative expansions of the homological block, they must include the overall prefactor.
Such prefactor takes a form of $e^{\frac{\kappa(\log x, \log a)}{\hbar}}$ where $\kappa$ is a polynomial with at most degree 2.
This prefactor can be fixed by using quantum $\widehat{A}$ and $\widehat{B}$ polynomials, which annihilate the homological block $F_K(x,a,q)$ \cite{EGGKPSS}.
We denote by $F'_{K}$ the positive expansion of homological block that contains the prefactor, and by $F_K$ the case that doesn't include the prefactor. 
In this notation, the balanced expansion $\mathcal{F}_K$ for $G_\mathbb{C}=SL(2,\mathbb{C})$ is expressed as
\begin{align}
\mathcal{F}_K(x,q) = \frac{1}{2} (F'_K(x,q) + F'_K(q^{-2}x^{-1},q))	\,	.
\label{fbhbr0}
\end{align}
For the fully unreduced normalization, we have
\begin{align}
\overline{\mathcal{F}}_K(x,q) = ((qx)^\frac{1}{2}-(qx)^{-\frac{1}{2}}) \mathcal{F}_K(x,q)	\,	.
\label{fbhb}
\end{align}

As in the case of the knot conormal, the homological block $F_K$ for a knot complement with a positive (negative) expansion admit the quiver form \eqref{gendt} \cite{Kucharski-quiver, EGGKPSS}.
Thus, by using the dictionary in section \ref{ssec:qfdbc}, we can obtain the 3d $\mathcal{N}=2$ theory whose half-index is the positive expansion $F_K$ of homological block.

%%%%%%%%%%%%%%%%%%%%%%%%%%%%%%%%%%%%%%%%%%%%%%%%%%%%%%%%%%%%%%%%%%%%%%%%%%%%%%%%%%%%

\subsubsection{3d $\mathcal{N}=2$ theory for $a$-deformed homological block $F_K$}
\label{sssec:afk}

As the $a$-deformation appears after the geometric transition, it is expected that the $a$-deformed homological block $F_K(x,a,q)$ is understood in the context of \eqref{mconf2} with a single M5 brane on $L_K$.
The variable $y$ that entered in the generating function of the HOMFLY polynomial is the eigenvalue of the holonomy along the longitude of a knot $K$ or the $S^1$ of the knot conormal $L_K$, which is topologically $S^1 \times \mathbb{R}^2$.
Meanwhile, the variable $x$ in $F_K(x,a,q)$ is the eigenvalue of the holonomy along the meridian of a knot $K$.
Therefore, it would be natural to associate $F_K(x,a,q)$ with the configuration that meridian is non-contractable.
A non-contractable meridian can be realized in a knot complement, and this can also be realized after the large $N$ transition \cite{AENV, EGGKPS}.

In the brane configuration \eqref{mconf1} with $M_3=S^3$, $N^{\prime}$ M5$'$ branes on $L_K$ with $N^{\prime} \sim \mathcal{O}(1)$ can be connected to $N$ M5 branes on $S^3$.
This gives $N^{\prime}$ M5$'$ branes on the knot complement $S^3 \backslash K$ and $N-N^{\prime}$ M5 branes on $S^3$.
Taking M5$'$ branes on $S^3 \backslash K$ to infinity and performing the large $N$ transition gives a Lagrangian filling $M_K$ of $\mathbb{R}\times \Lambda_K$ in $X$ whose classical asymptotic is $y \sim 1$ where $\Lambda_K$ is a Legendrian torus at infinity.
Thus the brane configuration is given by \eqref{mconf2} with $M_K$ instead of $L_K$ with a different K\"ahler parameter $\tilde{a} = q^{N-N^{\prime}}$.
We take $N^{\prime}=1$ here, so $\tilde{a}=q^{N-1}=q^{-1}a$.

We may ask whether the $a$-deformed homological block for a knot complement in the positive expansion is appropriate quantity to consider given the brane configuration with $M_K$.
There is also the balanced expansion for the $a$-deformed homological block, which has a symmetry $\mathcal{F}_K(x,a,q) = \mathcal{F}_K(a^{-1}x^{-1},a,q)$ \cite{EGGKPS}.
Considering the specialization $a=q^N$, the resulting $SL(N,\mathbb{C})$ homological block $\mathcal{F}^{SL(N,\mathbb{C})}(x,q)$ in the balanced expansion should have such property, because it is the $G_{\mathbb{C}}=SL(N,\mathbb{C})$ Chern-Simons partition function on a knot complement.
Meanwhile, the brane configuration under consideration has a single M5$'$ brane on $M_K$ or $L_K$, so we consider the $GL(1,\mathbb{C})$ Chern-Simons theory where the holonomy along the meridian (and longitude) is a complex parameter $\mathbb{C}^{*}$ classically \cite{Aganagic-Vafa-Q, AENV}, and there is no Weyl group action to consider in this case.
So it is expected that the $GL(1,\mathbb{C})$ Chern-Simons partition function on $M_K$ gives the $a$-deformed homological block $F_K(x,a,q)$ in the positive expansion.
Considering that a new K\"ahler parameter $\tilde{a}=q^{-1}a$ is associated with the configuration with $M_K$, we denote the 3d $\mathcal{N}=2$ theory obtained from $F_K(x,q\tilde{a},q)$ by $T[M^{(\tilde{a})}_K]$.	\\

Below, we provide a few examples from the standard quantity $F_K(x,a,q)$, which is a positive expansion without the prefactor, and then $T[M^{(\tilde{a})}_K]$ from $F_K(x,q\tilde{a},q)$. 
There is one difference between the $T[M^{(\tilde{a})}_K]$ obtained from $F_K(x,q\tilde{a},q)$ and the 3d $\mathcal{N}=2$ theory obtained from $F_K(x,a,q)$, and $T[M^{(\tilde{a})}_K]$ can be obtained from the latter.
The UV mixed CS level $k'_{gR}$ of $T[M^{(\tilde{a})}_K]$ between the gauge symmetries $U(1)$'s and $U(1)_R$ is given by $k'_{gR}=2k_{ga}+k_{gR}$ where $k_{ga}$ and $k_{gR}$ are the UV mixed CS levels between $U(1)$'s and $U(1)_a$, and between $U(1)$'s and $U(1)_R$ of the 3d $\mathcal{N}=2$ theory obtained from $F_K(x,a,q)$, respectively.
The rest is the same, while the notation changes from $a$ to $\tilde{a}$.	\\

Before providing some examples, we note that in this paper we consider the homological block $F_K$ for the abelian flat connection or the abelian branch.
There are also homological blocks $F_K$ or two-variable series invariants with possibly additional deformation variables for non-abelian branches and these also admit the quiver form \cite{EGGKPSS}.
By using the dictionary in section \ref{ssec:qfdbc}, we can obtain the 3d $\mathcal{N}=2$ theory whose half-index is the homological block for the non-abelian branch.
It would be interesting to study relations between the 3d $\mathcal{N}=2$ theories corresponding to the abelian branches and the non-abelian branches.

%%%%%%%%%%%%%%%%%%%%%%%%%%%%%%%%%%%%%%%%%%%%%%%%%%%%%%%%%%%%%%%%%%%%%%%%%%%%%%%%%%%%
\subsubsection*{\textbf{Unknot $0_1$}}

From \eqref{unrfk01}, the quiver form of the reduced homological block $F_K$ for the trefoil knot \cite{Kucharski-quiver, EGGKPSS} is given by
\begin{align}
F_{0_1}^{u.r.}(x,a,q) = \sum_{\mathbf{d} \geq 0} q^{\frac{1}{2}  \sum_{i,j=1}^{2} C_{ij}d_i d_j} \left( \prod_{j=1}^{2} \frac{1}{(q;q)_{d_j}} \right)
x^{d_1+d_2} a^{d_2} (-q^{\frac{1}{2}})^{-d_1}	
\end{align}
where
\begin{equation}
C_{ij} = 
\begin{pmatrix}
1	&0	\\
0	&0
\end{pmatrix}	\,	.
\end{equation}
From this, we have
\begin{itemize}[leftmargin=5mm]
\item $U(1)^2$ vector multiplets with the $\mathcal{D}$ boundary condition
\item 2 chiral multiplets $\Phi_i$, $i=1,2$ of R-charge 0 and charge $-\delta_{ij}$ under $U(1)_j$, $j=1,2$ with the D$_{c}$ boundary condition
\item UV (mixed) CS levels for $U(1)^2$ $-$ $U(1)^2$ : $
\begin{pmatrix}
\frac{1}{2}	&0	\\
0		&-\frac{1}{2}
\end{pmatrix}
$
\item UV mixed CS levels for $U(1)^2$ $-$ $U(1)_x$, $U(1)_a$, and $U(1)_R$
\begin{align*}
\left(1,1\right)	\,	,	\quad	\left(0,1\right)	\,	,	\quad	\left(\frac{1}{2},-\frac{1}{2}\right)	\,	.
\end{align*}
\end{itemize}

The $T[M^{(\tilde{a})}_K]$ obtained from $F_{0_1}^{u.r.}(x,q\tilde{a},q)$ is given by the same data as above but the UV mixed CS level for $U(1)^2$ $-$ $U(1)_R$ is $\big(\frac{1}{2}, \frac{3}{2}\big)$ and $U(1)_a \rightarrow U(1)_{\tilde{a}}$.

%%%%%%%%%%%%%%%%%%%%%%%%%%%%%%%%%%%%%%%%%%%%%%%%%%%%%%%%%%%%%%%%%%%%%%%%%%%%%%%%%%%%
\subsubsection*{\textbf{Trefoil $3_1^l$ (left-handed)}}

The reduced homological block for the trefoil knot \cite{Kucharski-quiver, EGGKPSS} is given by
\begin{align}
F_{3_1^{l}}(x,a,q)  = \sum_{k=0}^{\infty} (xq)^k \frac{(x;q^{-1})_k (aq^{-1};q)_k}{(q;q)_k}	\,	.
\end{align}
By using identities for $q$-series, the homological block can be expressed as
\begin{align}
F_{3_1}(x,a,q) = \sum_{\mathbf{d} \geq 0} q^{\frac{1}{2} \sum_{i,j=1}^{4} C_{ij}d_i d_j} \left( \prod_{j=1}^{4} \frac{1}{(q;q)_{d_j}} \right)
x^{2d_1 + 2d_2 + d_3 + d_4}
a^{d_1+d_3}
(-q^{\frac{1}{2}})^{3d_2 - d_3 + 2d_4}
\end{align}
where
\begin{equation}
C_{ij} = 
\begin{pmatrix}
0	&-1	&0	&-1	\\
-1	&-1	&0	&-1	\\
0	&0	&1	&0	\\
-1	&-1	&0	&0
\end{pmatrix}	\,	.
\end{equation}
The prefactor in this case is $qx$.

From the quiver form, the 3d $\mathcal{N}=2$ theory is given by
\begin{itemize}[leftmargin=5mm]
\item $U(1)^4$ vector multiplets with the $\mathcal{D}$ boundary condition
\item 4 chiral multiplets $\Phi_i$, $i=1, \ldots, 4$ of R-charge 0 and charge $-\delta_{ij}$ under $U(1)_j$, $j=1,\ldots, 4$ with the D$_c$ boundary condition
\item UV (mixed) CS level for $U(1)^4$ $-$ $U(1)^4$ : 
$
\begin{pmatrix}
-\frac{1}{2}	&-1			&0	&-1	\\
-1			&-\frac{3}{2}	&0	&-1	\\
0			&0			&1	&0	\\
-1			&-1			&0	&-\frac{1}{2}
\end{pmatrix}
$
\item UV mixed CS levels for $U(1)^4$ $-$ $U(1)_x$, $U(1)_a$, and $U(1)_R$
\begin{align*}
\left(2,2,1,1\right)	\,	,	\quad
\left(1,0,1,0\right)	\,	,	\quad
\left(-\frac{1}{2},\frac{5}{2},-\frac{3}{2},\frac{3}{2}\right)	\,	.
\end{align*}
\end{itemize}

The $T[M^{(a)}_K]$ from $F_{3_1^{l}}(x,q\tilde{a},q)$ is given by the same data as above except that the UV mixed CS level for $U(1)^4$ $-$ $U(1)_R$ is $\left(\frac{3}{2},\frac{5}{2},\frac{1}{2},\frac{3}{2}\right)$ and $U(1)_a \rightarrow U(1)_{\tilde{a}}$.

%%%%%%%%%%%%%%%%%%%%%%%%%%%%%%%%%%%%%%%%%%%%%%%%%%%%%%%%%%%%%%%%%%%%%%%%%%%%%%%%%%%%
\subsubsection*{\textbf{Trefoil $3^r_1$ (right-handed)}}

The reduced homological block for right-handed trefoil knot \cite{EGGKPSS} is
\begin{align}
F_{3_1^r}(a,x,q) = \sum_{0 \leq \mathbf{d}} q^{\frac{1}{2} \sum_{i,j=1}^{4} C_{ij}d_i d_j} \left( \prod_{j=1}^{4} \frac{1}{(q;q)_{d_j}} \right)
x^{d_1 + d_2 + d_3 + d_4}
a^{d_2 + d_3 + d_4}
(-q^{\frac{1}{2}})^{2d_1 - d_3 - d_4}
\end{align}
where
\begin{equation}
C_{ij} = 
\begin{pmatrix}
0	&1	&0	&0	\\
1	&0	&1	&0	\\
0	&1	&1	&0	\\
0	&0	&0	&1
\end{pmatrix}	\,	.
\end{equation}
The prefactor of $F_{3_1^r}(x,q)$ is $q^2 x$.

The corresponding 3d $\mathcal{N}=2$ theory is 
\begin{itemize}[leftmargin=5mm]
\item $U(1)^4$ vector multiplets with the $\mathcal{D}$ boundary condition
\item 4 chiral multiplets $\Phi_i$, $i=1, \ldots, 4$ of R-charge 0 and charge $-\delta_{ij}$ under $U(1)_j$, $j=1, \ldots, 4$ with the D$_{c}$ boundary condition
\item UV (mixed) CS level for $U(1)^4$ $-$ $U(1)^4$ : 
$
\begin{pmatrix}
-\frac{1}{2}	&1			&0			&0	\\
1			&-\frac{1}{2}	&1			&0	\\
0			&1			&\frac{1}{2}	&0	\\
0			&0			&0			&\frac{1}{2}
\end{pmatrix}
$
\item UV mixed CS levels for $U(1)^4$ $-$ $U(1)_x$, $U(1)_a$, and $U(1)_R$
\begin{align*}
\left(1,1,1,1\right)	\,	,	\quad
\left(0,1,1,1\right)	\,	,	\quad
\left(\frac{3}{2},-\frac{1}{2},-\frac{3}{2},-\frac{3}{2}\right)	\,	.
\end{align*}
\end{itemize}

The $T[M^{(\tilde{a})}_K]$ for $F_{3_1^r}(x,q\tilde{a},q)$ has the same data as above but the mixed UV CS level for $U(1)^4$ $-$ $U(1)_R$ is $\left(\frac{3}{2},\frac{3}{2},\frac{1}{2},\frac{1}{2}\right)$ and $U(1)_a \rightarrow U(1)_{\tilde{a}}$.

%%%%%%%%%%%%%%%%%%%%%%%%%%%%%%%%%%%%%%%%%%%%%%%%%%%%%%%%%%%%%%%%%%%%%%%%%%%%%%%%%%%%
\subsubsection*{\textbf{Figure-eight knot $4_1$}}

The reduced homological block for the figure-eight knot \cite{EGGKPSS} is given by
\begin{align}
F_{4_1}(x,a,q) = \sum_{\mathbf{d} \geq 0} q^{\frac{1}{2} \sum_{i,j=1}^{6} C_{ij}d_i d_j} \left( \prod_{j=1}^{6} \frac{1}{(q;q)_{d_j}} \right)
x^{\sum_{i=1}^{6}d_i}
a^{\sum_{i=4}^{6}d_i}
(-q^{\frac{1}{2}})^{2\sum_{i=1}^{3}d_i-\sum_{i=4}^{6}d_i}
\label{afk41q}
\end{align}
where
\begin{equation}
C_{ij} = 
\begin{pmatrix}
0	&0	&0	&0	&0	&0	\\
0	&0	&-1	&-1	&0	&0	\\
0	&-1	&0	&0	&1	&0	\\
0	&-1	&0	&1	&1	&0	\\
0	&0	&1	&1	&1	&0	\\
0	&0	&0	&0	&0	&1	
\end{pmatrix}	\,	.
\end{equation}
The prefactor of $F_{4_1}(x,q)$ is $qx$.

Therefore, the 3d $\mathcal{N}=2$ theory for $F_{4_1}(x,a,q)$ is given by 
\begin{itemize}[leftmargin=5mm]
\item $U(1)^6$ vector multiplets with the $\mathcal{D}$ boundary condition
\item 6 chiral multiplets $\Phi_i$, $i=1, \ldots, 6$ of R-charge 0 and charge $-\delta_{ij}$ under $U(1)_j$, $j=1, \ldots, 6$ with the D$_{c}$ boundary condition
\item UV (mixed) CS levels for $U(1)^6$ $-$ $U(1)^6$ : 
$
\begin{pmatrix}
-\frac{1}{2}	&0	&0	&0	&0	&0	\\
0			&-\frac{1}{2}	&-1	&-1	&0	&0	\\
0			&-1	&-\frac{1}{2}	&0	&1	&0	\\
0			&-1	&0	&\frac{1}{2}	&1	&0	\\
0			&0	&1	&1	&\frac{1}{2}	&0	\\
0			&0	&0	&0	&0	&\frac{1}{2}	
\end{pmatrix}
$
\item UV mixed CS levels for $U(1)^6$ $-$ $U(1)_x$, $U(1)_a$, and $U(1)_R$: 
\begin{align*}
\left(1,1,1,1,1,1\right)	\,	,	\quad
\left(0,0,0,1,1,1\right)	\,	,	\quad
\left(\frac{3}{2},\frac{3}{2},\frac{3}{2},-\frac{3}{2},-\frac{3}{2},-\frac{3}{2}\right)	\,	.
\end{align*}
\end{itemize}

The theory $T[M^{(\tilde{a})}_K]$ obtained from $F_{4_1}(x,q\tilde{a},q)$ is given by the same data as above but the UV mixed CS level for $U(1)^6$ $-$ $U(1)_R$ is $\left(\frac{3}{2},\frac{3}{2},\frac{3}{2},\frac{1}{2},\frac{1}{2},\frac{1}{2}\right)$ and $U(1)_a \rightarrow U(1)_{\tilde{a}}$.

%%%%%%%%%%%%%%%%%%%%%%%%%%%%%%%%%%%%%%%%%%%%%%%%%%%%%%%%%%%%%%%%%%%%%%%%%%%%%%%%%%%%

\subsubsection{The case of $G=SU(2)$ and surgery}

The homological block $F^{SL(2,\mathbb{C})}_K(x, q)$ for $G_{\mathbb{C}}=SL(2,\mathbb{C})$ can be obtained by taking a specialization $a=q^2$ on the $a$-deformed homological block $F_K(x,a,q)$. 
Since the expression of the $a$-deformed homological block $F_K(x,a,q)$ used above is a positive expansion, we obtain the homological block $F^{SL(2,\mathbb{C})}_K(x, q)$ in the positive expansion.
In order to have a proper 3d-3d correspondence with the partition function of $G_{\mathbb{C}}=SL(2,\mathbb{C})$ Chern-Simons theory on $S^3 \backslash K$, one should have the balanced expression $\mathcal{F}^{SL(2,\mathbb{C})}_K(x,q)$, which is obtained by the average of the positive and the negative expansions. 
However, it is not obvious to obtain a 3d $\mathcal{N}=2$ theory from such a resulting expression obtained from the average. 
But all information is basically encoded in the positive (or negative) expansion, or in other words, the balanced expansion is completely determined by the positive (or negative) expansion, so we may consider a 3d $\mathcal{N}=2$ theory whose half-index is the positive expansion $F^{SL(2,\mathbb{C})}_K(x,q)$ of homological block without the prefactor.
We call such theory $T^{p}[S^3 \backslash K]$ where $p$ denotes the positive expansion.
We omit $SL(2,\mathbb{C})$ in $F^{SL(2,\mathbb{C})}_K(x,q)$ and simply use the notation $F_K(x,q)$.	\\

The theory $T^{p}[S^3 \backslash K]$ can be readily obtained by using information of previous section.
For example, for the figure-eight knot, with $a=q^2$ \eqref{afk41q} becomes 
\begin{align}
F_{4_1}(x,q) = \sum_{\mathbf{d} \geq 0} q^{\frac{1}{2} \sum_{i,j=1}^{6} C_{ij}d_i d_j} \left( \prod_{j=1}^{6} \frac{1}{(q;q)_{d_j}} \right)
x^{\sum_{i=1}^{6} d_i}
(-q^{\frac{1}{2}})^{2\sum_{i=1}^{3} d_i + 3\sum_{i=4}^{6} d_i}	\,	.
\label{afk41qsl2}
\end{align}
Therefore, other information of the theory are the same except that we have no $U(1)_a$ and the UV mixed CS levels between $U(1)^6$ and $U(1)_R$ becomes $(\frac{3}{2}, \frac{3}{2}, \frac{3}{2}, \frac{5}{2}, \frac{5}{2}, \frac{5}{2})$.
In general, given the quiver form \eqref{gendt} or \eqref{gendtn}, since the dependence of $a$ enters only via $x_j$ in \eqref{gendtn}, other information stays the same but due to $a=q^2$ there is no $U(1)_a$ symmetry and the UV mixed CS level $k'_{gR}$ of $T^{p}[S^3 \backslash K]$ between gauge symmetries $U(1)$'s and $U(1)_R$ symmetry is given by $k'_{gR} = 4k_{ga} + k_{gR}$ where $k_{ga}$ and $k_{gR}$ are the UV mixed CS levels between $U(1)$'s and $U(1)_a$, and between $U(1)$'s and $U(1)_R$ of the 3d $\mathcal{N}=2$ theory obtained from $F_K(x,a,q)$, respectively.	\\

Though physical meaning of the 3d $\mathcal{N}=2$ theory $T^{p}[S^3 \backslash K]$ itself is not clear, it can be used when considering the surgery of 3-manifolds.
Let $\mu$ and $\lambda$ be the meridian and the longitude of the knot complement $S^3 \backslash K$. 
The $p/h$ surgery $S^3_{p/h}(K)$ can be obtained by gluing $S^3 \backslash K$ and the solid torus $D^2 \times S^1$,
\begin{align}
S^3_{p/h}(K) = S^3 \backslash K \, \cup \, (D^2 \times S^1)
\end{align}
where $p \mu + h \lambda$ is identified with the meridian $\partial D^2 \times pt$ of the solid torus $D^2 \times S^1$ and $p$ and $h$ are coprime.
A surgery formula for a knot complement $S^3 \backslash K$ with the surgery coefficient $p/h$ is given by \cite{Gukov-Manolescu}
\begin{align}
\widehat{Z}_b(S^3_{p/h}(K)) \simeq \oint_{|z|=1} \frac{dz}{2 \pi i z} (z-z^{-1}) \, \mathcal{F}_{K}(q^{-1}z^2,q) \, \sum_{n \in \mathbb{Z}} \widehat{Z}_b(\mathbb{S}_{p/h};z,n,q)
\label{sform}
\end{align}
up to an overall factor that depends on sign and $q$ where $b$ is Spin$^c$ structure.\footnote{The formula works for $p$ and $h$ in certain ranges, not for all of $p$ and $h$ \cite{Gukov-Manolescu}.}
Here, $\mathcal{F}_{K}(q^{-1}x,q)$ denotes the reduced homological block \eqref{fbhbr0} in the balanced expansion for a knot complement $S^3 \backslash K$.
But since the convention in \cite{Gukov-Manolescu} is different from the convention here, $x$ in $\mathcal{F}_{K}(x,q)$ should be replaced by $q^{-1}x$.\footnote{In the convention used in \cite{Gukov-Manolescu}, $x_{\text{GM}}=q^n$ for the $n$-dimensional representation of $G=SU(2)$, while $x_\text{here}=q^r=q^{n-1}=q^{-1}x_{\text{GM}}$ for the $n-1$-dimensional representation.}	\\

$\widehat{Z}_b(\mathbb{S}_{p/h};z,n,q)$ is a homological block of the solid torus $\mathbb{S}_{p/h}$ with the surgery information encoded.
The homological block $\widehat{Z}_b(\mathbb{S}_{p/h};z,n,q)$ can be obtained from the homological block of the plumbed 3-manifold where one of nodes is ungauged.	\\

We briefly review the homological block $\widehat{Z}_b(\mathbb{S}_{p/h};z,n,q)$ for the solid torus \cite{Gukov-Manolescu}.
Consider a plumbing graph $\Gamma$ with vertices $v \in \text{Vert}$ that are connected by edges.
Given a vertex $v$, the degree $\text{deg}(v)$ denotes the number of edges meeting at a vertex $v$.
The weight of a vertex $v$ is denoted by $m_v$.
Then for the plumbing graph with $s$ vertices, the adjacency matrix $M$ is an $s$ by $s$ matrix and is given by
\begin{align}
M_{v_1 v_2} = 
\begin{cases}	
\ m_v	&	\qquad	v=v_1=v_2	\\
\ 1		&	\qquad	\text{$v_1$ and $v_2$ connected}	\\
\ 0		&	\qquad	\text{otherwise}
\end{cases}	\,	.
\end{align}
The homological block of the plumbed 3-manifold $Y_{\Gamma}$ is given by the contour integral
\begin{align}
\widehat{Z}_b(q) \simeq \text{v.p.} \oint_{|z_v|=1} \prod_{v \in \text{Vert}} \frac{dz_v}{2 \pi i z_v} \bigg( z_v - \frac{1}{z_v} \bigg)^{2-\text{deg}(v)} \Theta_b^{-M}(\vec{z})
\label{pzhat}
\end{align}
up to an overall factor that depends on sign and $q$.
The homological block $\widehat{Z}_b(q)$ is labelled by $b \in \text{Spin}^c(Y_{\Gamma})$ where Spin$^c$ structures are affinely isomorphic to abelian flat connections.
Here, v.p. means the principal value integral, and
\begin{align}
\Theta_{b}^{-M}(\vec{z}) = \sum_{\vec{l} \in 2 M \mathbb{Z}^s+ \vec{b}} q^{- \frac{(\vec{l}, M^{-1}\vec{l})}{4}} \prod_{v \in \text{Vert}} z_v^{l_v}	\,	.
\label{ptheta}
\end{align}
The formula \eqref{pzhat} is well-defined for the weakly negative definite condition on the adjacency matrix $M$ \cite{Gukov-Manolescu}.	\\

The homological block for a plumbed knot complement $Y_{\Gamma, v_0}$ can be obtained by ``ungauging" a vertex $v_0$ of the plumbing graph, 
\begin{align}
\widehat{Z}_{b}(Y;z,n,q) \simeq \bigg( z- \frac{1}{z} \bigg)^{1-\deg(v_0)} \oint_{|z_v|=1} \prod_{\stackrel{v \in \text{Vert}}{v \neq v_0}} \frac{dz_v}{2\pi i z_v} \bigg( z_v - \frac{1}{z_v} \bigg)^{2- \deg(v)} \Theta_{b}^{-M}(\vec{z})
\label{pzhat1}
\end{align}
where $b$ denotes the relative Spin$^{c}$ structure $\text{Spin}^{c}(Y_{\Gamma, v_0}, \partial Y_{\Gamma, v_0})$ and the integer $n \in \mathbb{Z}$ for the vertex $v_0$ in the entries of $\vec{n}\in \mathbb{Z}^s$ in $\Theta_{b}^{-M}(\vec{z}) = \sum_{\vec{l} \in 2 M \vec{n}+ \vec{b}} q^{- \frac{(\vec{l}, M^{-1}\vec{l})}{4}} \prod_{v \in \text{Vert}} z_v^{l_v}$ is fixed.	\\

The homological block $\widehat{Z}_{b}(\mathbb{S}_{p/h};z,n,q)$ for the solid torus $\mathbb{S}_{p/h}$ can be obtained from a linear plumbing graph of $s$ nodes whose weights are $k_1, k_2, \ldots, k_s$ such that
\begin{align}
\frac{p}{h} = k_1 - \cfrac{1}{k_2 -\cfrac{1}{\ddots -\cfrac{1}{k_s}}}
\end{align}
with the first node whose weight is $k_1$ ungauged.
The adjacency matrix $M$ of the linear plumbing graph is given by 
\begin{align}
M=
\begin{pmatrix}
k_1	&1	&	&	&	&	\\
1	&k_2	&1	&	&	&	\\
	&1	&k_3	&	&	&	\\
	&	&	&\ddots	&	&	\\	
	&	&	&	&k_{s-1}	&1	\\	
	&	&	&	&1	&k_s
\end{pmatrix}	\,	.
\end{align}

\vspace{3mm}

Under the conjugation $b \leftrightarrow -b$, the homological blocks for plumbed 3-manifolds are invariant $\widehat{Z}_{b}(S^3_{p/h}(K)) = \widehat{Z}_{-b}(S^3_{p/h}(K))$, and for plumbed knot complement $Y_{\Gamma, v_0}$, $\widehat{Z}_{b}(Y_{\Gamma, v_0};z,n,q) = - \widehat{Z}_{-b}(Y_{\Gamma, v_0};z^{-1},-n,q)$ \cite{Gukov-Manolescu}.
Therefore, by using \eqref{fbhb}, the surgery formula \eqref{sform} can be expressed as
\begin{align}
\widehat{Z}_b(S^3_{p/h}(K)) \simeq \frac{1}{2} \oint_{|z|=1} \frac{dz}{2 \pi i z} (z-z^{-1}) F'_{K}(q^{-1}z^2,q)  \bigg( \sum_{n \in \mathbb{Z}} \widehat{Z}_{b}(\mathbb{S}_{p/h};z,n,q) + \widehat{Z}_{-b}(\mathbb{S}_{p/h};z,n,q) \bigg)	\,	.
\label{sform1}
\end{align}
Here, $F'_K(x,q)$ includes the prefactor $e^{\frac{1}{\hbar}\kappa(\log x, \log a)}$ with $a=q^2$.	\\

Since we know the theory for $F_{K}(q^{-1}z^2,q)$, we may try to interpret the gluing formula of the $p/h$ surgery in the context of 3d $\mathcal{N}=2$ theory, which gives $T[S^3_{p/r}(K)]$.
Though it is expected that there is a formulation of \eqref{sform} or \eqref{sform1} in the language of $G=SU(2)$, we interpret the formula in terms of an abelian gauge theory by considering the contour integral for $z$ in \eqref{sform1} as the half-index of a $U(1)_z$ vector multiplet in the Neumann boundary condition that couples to 3d $\mathcal{N}=2$ theories corresponding to the factors in the integrand. 	\\

For the case that the resulting closed 3-manifold has more than one Spin$^{c}$ structure (or one abelian flat connection), a sum of two theta functions arises as in \eqref{sform1}, so we focus on a simpler case that there is a single abelian flat connection where there is just one theta function.
For this case, we take $|p|=1$, and the abelian flat connection is identified with $b \equiv 0$ or $b \equiv \frac{1}{2}$ mod $\mathbb{Z}$, when $r$ is odd and even, respectively.
For convenience, we denote both of them simply by 0.
For such $b$, the gluing formula \eqref{sform1} can be expressed as
\begin{align}
\widehat{Z}_0(S^3_{-1/h}(K)) \simeq \oint_{|z|=1} \frac{dz}{2 \pi i z} (z-z^{-1}) F'_{K}(q^{-1}z^2,q)  \sum_{n \in \mathbb{Z}} \widehat{Z}_0(\mathbb{S}_{-1/h};z,n,q)
\label{sform2}
\end{align}
where we took $-1/h$.	\\

Given the surgery formula \eqref{sform2}, we discuss 3d $\mathcal{N}=2$ theories for each factors.	
Regarding $\sum_{n \in \mathbb{Z}} \widehat{Z}_0(\mathbb{S}_{-1/h};z,n,q)$, we need to interpret the formula \eqref{pzhat} and \eqref{pzhat1}.\footnote{There are some studies on 3d $\mathcal{N}=2$ theories for plumbing graphs in \cite{Gukov-Putrov-Vafa, EKSW, Chung-indexplumbing, Chun-Gukov-Park-Sopenko}.}
For a single vertex with weight $-p$, the 3d $\mathcal{N}=2$ theory for the vertex is known as a $G=SU(2)$ vector multiplet with the Neumann boundary condition, Chern-Simons level $p$, the Neumann boundary condition for an adjoint chiral multiplet with R-charge 2, and the boundary degree of freedom that is captured in the theta function $\Theta_a(z) = \prod_{l \in p \mathbb{Z}+a} q^{\frac{l^2}{p}} z^{2l}$ which cancels anomaly from the Chern-Simons term with the level $p$ \cite{Gukov-Putrov-Vafa,Gukov-Pei-Putrov-Vafa}.

For the edge, we may try to interpret and engineer it in terms of a $U(1)$ theory.
The factor $z-z^{-1}$ in \eqref{pzhat} or \eqref{pzhat1} can be expressed as
\begin{align}
z-\frac{1}{z} = z (1-z^{-2}) = - \frac{(z;q)_\infty}{(qz;q)_\infty} \frac{(qz^{-1};q)_\infty}{(z^{-1};q)_\infty} \frac{(z^{-2};q)_\infty}{(qz^{-2};q)_\infty}	\,	.
\label{zzinv}
\end{align}
Therefore, up to an overall sign, we have a $U(1)$ theory for the $z-z^{-1}$ factor,
\begin{itemize}[leftmargin=5mm]
\item 3 chiral multiplets charged $(-1,2)$, $(1,0)$, $(2,2)$ under $U(1)_z \times U(1)_R$ with the D boundary condition
\item 3 chiral multiplets charged $(1,2)$, $(-1,0)$, $(-2,2)$ under $U(1)_z \times U(1)_R$ with the N boundary condition
\end{itemize}
and we denote this theory by $T_{v}^{(1/2)}$ or $T^{(1/2)}$ for convenience where $(1/2)$ means the half the contribution to the vertex.\footnote{We note that the first two factors in the numerator and in the denominator of \eqref{zzinv} form theta functions \eqref{jtheta} and can be regarded as contributions of the 2d $\mathcal{N}=(0,2)$ Fermi and chiral multiplets charged $(-1,1)$ and $(1,2)$ under $U(1)_z \times U(1)_R$, respectively.}
In order to have the contribution $(z-z^{-1})^2$ for the vertex $v$, we have two copies of $T_{v}^{(1/2)}$.
We can see that there is no anomaly at all for $T_{v}^{(1/2)}$.

The edge connecting two vertices, $v_1$ and $v_2$, have the contribution $(z_{v_1}-z_{v_1}^{-1})^{-1}(z_{v_2}-z_{v_2}^{-1})^{-1}$.
The factor $(z-z^{-1})^{-1}$ can be realized by the matter contents
\begin{itemize}[leftmargin=5mm]
\item 3 chiral multiplets charged $(-1,0)$, $(1,2)$, $(2,0)$ under $U(1)_z \times U(1)_R$ with the D boundary condition
\item 3 chiral multiplets charged $(1,0)$, $(-1,2)$, $(-2,0)$ under $U(1)_z \times U(1)_R$ with the N boundary condition
\end{itemize}
and we denote this theory by $T_{v}^{(-1/2)}$.\footnote{Similarly, instead of the first two chiral multiplets for each, we may have the 2d $\mathcal{N}=(0,2)$ Fermi and chiral multiplets charged $(1,1)$ and $(-1,2)$ under $U(1)_z \times U(1)_R$, respectively.}
Therefore, the theory $T_{e_{v_1 v_2}}$ for the edge between the vertices $v_1$ and $v_2$ can be given by the product $T_{v_1}^{(-1/2)} T_{v_2}^{(-1/2)}$.
Similarly, this theory is also anomaly-free.

The adjacency matrix with minus sign $-M$ corresponds to the level of the mixed $G=SU(2)$ Chern-Simons term between the vector multiplets at the vertices of the plumbing graph.
Since we discuss in terms of $U(1)$'s, the mixed Chern-Simons levels between the vector multiplets $U(1)_{v_1}$ and $U(1)_{v_2}$ are given by $-2M_{v_1 v_2}$.
The theta function $\Theta_{b}^{-M}(\vec{z})$ depends on the boundary condition and it cancels the anomaly from the mixed Chern-Simons terms with the levels $-2M$.
The precise 2d $\mathcal{N}=(0,2)$ boundary theory that gives $\Theta_{b}^{-M}(\vec{z})$ is not known, and in this paper we just say that there is a conformal field theory at the boundary whose character is $\Theta_{b}^{-M}(\vec{z})$ for a given boundary condition $b$.	\\

For the solid torus $\mathbb{S}_{-1/h}$, which is described by the linear plumbing graph, $k=(k_1, \ldots, k_s)$, the factor $z-z^{-1}$ appears only at the vertex $v_s$ of the quiver, 
\begin{align}
\sum_{n \in \mathbb{Z}} \widehat{Z}_0(\mathbb{S}_{-1/h};z,n,q) \simeq \oint_{|z_v|=1} \prod_{v \in \{2, \ldots, s \}} \frac{dz_v}{2\pi i z_v}  \bigg( z_{v_s} - \frac{1}{z_{v_s}} \bigg) \Theta_{0}^{-M}(\vec{z})
\label{st1r}
\end{align}
up to an overall factor.
$-1/h$ can be realized, for example, by a choice $k=(-1, \underbrace{-2, \ldots, -2}_{h-1})$ of the linear plumbing graph.
With the sum over $n \in \mathbb{Z}$, there is no fixed component in $\vec{n} \in \mathbb{Z}^s$ of $\Theta_{0}^{-M}(\vec{z})$ in \eqref{st1r}.
For the theta function part, there would be a boundary CFT whose character of the module is $\Theta_{0}^{-M}(\vec{z})$.
Then, the 3d $\mathcal{N}=2$ theory that describes \eqref{st1r} is a linear quiver theory where
\begin{itemize}[leftmargin=5mm]
\item $U(1)$ vector multiplet for each node $v_1, \ldots, v_s$
\item Chern-Simons coupling between $U(1)^s$ and $U(1)^s$ whose level is $-2M$
\item the theory $T^{(1/2)}_{v_s}$ at the node $v_s$
\item boundary CFT whose character is $\Theta_{0}^{-M}(\vec{z})$ in \eqref{st1r}
\end{itemize}
and we call this theory $T[\mathbb{S}_{-1/h}]$.	\\

For a knot complement $S^3 \backslash K$, the $F'_K(q^{-1}z^2, q)$ is given by a product of the prefactor and $F_K(q^{-1}z^2, q)$.
The theory for $F_K(q^{-1}z^2, q)$ is obtained from the theory $T^{p}[S^3 \backslash K]$.
$U(1)_x$ is replaced with $U(1)_z$, and the UV mixed CS levels $k_{jz}$ between $U(1)_j$, $j=1, \ldots, L$ gauge symmetries and $U(1)_z$ is given by $k_{jz} = 2 k_{jx}$ where $k_{jx}$ is the level for the mixed CS coupling between $U(1)_j$ and $U(1)_x$ of $T^{p}[S^3 \backslash K]$.
Also, the UV mixed CS levels between $U(1)_j$, $j=1, \ldots, L$ and $U(1)_R$ is given by  $k_{jR}-2k_{jx}$, $j=1, \ldots, L$, where $k_{jR}$ is the UV mixed CS level between $U(1)_j$, $j=1, \ldots, L$ and $U(1)_R$ of $T^{p}[S^3 \backslash K]$.
We call this theory $T^{p2}[S^3 \backslash K]$.

The overall prefactor depends on the knot under consideration.
For example, for the figure-eight knot $K=4_1$, the overall prefactor is given by $qx=x_{GM}$.
Thus, this contributes in $F'_K(q^{-1}z^2, q)$ by $z^2$, which can be realized as
\begin{align}
\left(- \frac{(z;q)_\infty}{(qz;q)_\infty} \frac{(qz^{-1};q)_\infty}{(z^{-1};q)_\infty}\right)^2	\,	,
\end{align}
\textit{c.f.} \eqref{zzinv}.
This can be obtained as the half-index of the theory
\begin{itemize}[leftmargin=5mm]
\item 2 chiral multiplets charged $(-1,2)$, $(1,0)$ under $U(1)_z \times U(1)_R$ with the D boundary condition
\item 2 chiral multiplets charged $(1,2)$, $(-1,0)$ under $U(1)_z \times U(1)_R$ with the N boundary condition.
\item UV mixed CS level for $U(1)_z$ and $U(1)_R$, $k_{zR}=4$
\end{itemize}
A monomial of $z$ with possibly some powers of $q$ can be engineered in this way, and we denote such theory for the prefactor by $T^{\text{p.f.}}[S^3 \backslash K]$.	\\

In addition, there is $(z-z^{-1})$ factor in the integrand of \eqref{sform2}.
This factor has already appeared in \eqref{zzinv}, so the theory for $(z-z^{-1})$ in the integrand of \eqref{sform2} is given by $T^{(1/2)}$.	\\

Therefore, from \eqref{sform2}, the 3d $\mathcal{N}=2$ theory $T[S^3_{-1/h}(K)]$ for closed 3-manifold $S^3_{-1/h}(K)$ obtained by the $-1/h$ surgery of a knot complement $S^3 \backslash K$ is given by
\begin{itemize}[leftmargin=5mm]
\item $U(1)_z$ vector multiplet with the Neumann boundary condition, which is coupled to $T^{p2}[S^3 \backslash K]$, $T^{\text{p.f.}}[S^3 \backslash K]$, $T^{(1/2)}$, and $T[\mathbb{S}_{-1/h}]$.
\end{itemize}

We note that the approach discussed here has a limitation that doesn't tell how to incorporate the refinement $U(1)_t$.
It is expected that a natural way to incorporate $U(1)_t$ would be available if an $SU(2)$ approach is understood.

%%%%%%%%%%%%%%%%%%%%%%%%%%%%%%%%%%%%%%%%%%%%%%%%%%%%%%%%%%%%%%%%%%%%%%%%%%%%%%%%%%%%
%%%%%%%%%%%%%%%%%%%%%%%%%%%%%%%%%%%%%%%%%%%%%%%%%%%%%%%%%%%%%%%%%%%%%%%%%%%%%%%%%%%%
%%%%%%%%%%%%%%%%%%%%%%%%%%%%%%%%%%%%%%%%%%%%%%%%%%%%%%%%%%%%%%%%%%%%%%%%%%%%%%%%%%%%

\subsection{3d $\mathcal{N}=2$ theory for closed 3-manifold and logarithmic CFT}
\label{ssec:lcft}

We are interested in the quiver form that appears in homological blocks for closed 3-manifolds and characters of conformal field theory.

If there is no contribution from the 3d bulk, the half-index of the 3d $\mathcal{N}=2$ theory is given by the character of a 2d rational CFT or the elliptic genus, which has modularity.
In the presence of 3d bulk degrees of freedom, the modularity is spoiled in a specific way, and the half-index with the contributions from the bulk exhibits the quantum modularity \cite{Zagier-quantum, CCFGH}. 
In addition, such half-index is expected to be identified with the character of the corresponding logarithmic conformal field theory (log-CFT) \cite{CCFGH, CCFFGHP}.

Meanwhile, it is known that the character of CFT admit so called the fermionic form \cite{Kedem-Klassen-McCoy-Melzer, Nahm-cft}, which takes a similar form with \eqref{gendt} or \eqref{gendtn}.
Thus, it is expected that given a 3-manifold $M_3$ we can obtain from the character of $\text{log-CFT}[M_3]$ the 3d $\mathcal{N}=2$ theory $T[M_3]$ with the 2d $\mathcal{N}=(0,2)$ boundary conditions, and vice versa.	\\

We consider Seifert manifolds with three singular fibers as examples.
The homological block for Seifert manifolds with three singular fibers is given by a linear combination with integer coefficients of the false theta function $\widetilde{\Psi}_{P}^{(l)}(q)$,
\begin{align}
\widetilde{\Psi}_{p}^{(l)}(q) = \sum_{n=0}^{\infty} \psi_{2p}^{(l)}(n) q^{\frac{n^2}{4p}}
\end{align}
where
\begin{align}
\psi_{2p}^{(l)}(n) =
\begin{cases}
\pm 1	&	\text{if $n \equiv \pm l$ mod $2p$}	\\
0		&	\text{otherwise}
\end{cases}	\,	.
\end{align}
It is known \cite{CCFGH, CCFFGHP} that the corresponding log-CFT for the Seifert manifold with three singular fibers is the $(1,p)$ singlet model \cite{Kausch-1991}.
The $(1,p)$ singlet algebra has modules $M_{1,s}$, $1\leq s \leq p$, and their characters are given by
\begin{align}
\chi_{M_{1,s}} = \frac{\widetilde{\Psi}_{p}^{(p-s)}(q)}{\eta(q)}
\end{align}
where $\eta(q)$ is the Dedekind eta function $\eta(q)=q^{\frac{1}{24}}(q;q)_\infty$.
For example, the homological block for Brieskorn sphere $\Sigma(2,3,5)$ is given by
\begin{align}
Z^{SL(2,\mathbb{C})}_{\Sigma(2,3,5)} \simeq 2 q^{\frac{1}{120}} - \widetilde{\Psi}_{30}^{(1)}(q) - \widetilde{\Psi}_{30}^{(11)}(q) - \widetilde{\Psi}_{30}^{(19)}(q) - \widetilde{\Psi}_{30}^{(29)}(q)	\,	.
\label{hbs235}
\end{align}
This corresponds to the character of the module of the $(1,p)$ singlet model with $p=30$ and $s=1,11,19,29$ \cite{CCFGH, CCFFGHP},
\begin{align}
Z^{SL(2,\mathbb{C})}_{\Sigma(2,3,5)} \simeq 2 q^{\frac{1}{120}} - \chi(M_{1,1} \oplus M_{1,11} \oplus M_{1,19} \oplus M_{1,29})	\,	.
\end{align}
Such correspondence also holds between Seifert manifolds with four singular fibers and the $(p,p')$ singlet models \cite{CCFFGHP}.	\\

We see that the homological block is expressed as the sum of the characters, and such expression seems not suitable for extracting the 3d $\mathcal{N}=2$ theory directly as we have done.\footnote{As a side remark, the character of the module of the $(1,p)$ triplet VOA that contains the $(1,p)$ singlet VOA as a subalgebra can be expressed as a fermionic form \cite{Feigin-Feigin-Tipunin}. 
However, we see that in general some of effective Chern-Simons levels in the fermionic form in \cite{Feigin-Feigin-Tipunin} are half-integers, so the fermionic form in \cite{Feigin-Feigin-Tipunin} itself seems not appropriate for the interpretation as the 3d $\mathcal{N}=2$ theories that is obtained from the quiver form \eqref{gendt} or \eqref{gendtn}.} 
However, the homological blocks that admit a quiver form as a whole are known for some closed 3-manifolds.
More precisely, we see that such quiver form is slightly different from the quiver form of \eqref{gendt}, but it is still possible to translate it to a 3d $\mathcal{N}=2$ theory.

For example, it is known \cite{CCFGH} that \eqref{hbs235} can be expressed as
\begin{align}
q^{\frac{1}{120}} \bigg( 2 - \sum_{k = 0}^{\infty} \frac{(-1)^{k} q^{\frac{3}{2}k^2 -\frac{1}{2}k}}{(q^{k+1};q)_k}  \bigg)	\,	.
\label{hbs2351}
\end{align}
The term $\sum_{k = 0}^{\infty} \frac{(-1)^{k} q^{\frac{3}{2}k^2 -\frac{1}{2}k}}{(q^{k+1};q)_k}$ in \eqref{hbs2351} can also be expressed as
\begin{align}
\sum_{k = 0}^{\infty} q^{\frac{3}{2}k^2} (-q^{\frac{1}{2}})^{-k} \frac{(q^{2k+1};q)_\infty}{(q^{k+1};q)_{\infty}} 	\,	.
\label{q235}
\end{align}
As mentioned above, this is not exactly the quiver form as in \eqref{gendt} or \eqref{gendtn}, but we can still obtain the 3d $\mathcal{N}=2$ theory whose half-index is \eqref{q235}, which is
\begin{itemize}[leftmargin=5mm]
\item $U(1)_g$ vector multiplet with the $\mathcal{D}$ boundary condition
\item 1 chiral multiplet charged $(1,2)$ under $U(1)_g \times U(1)_R$ with the N boundary condition
\item 1 chiral multiplet charged $(-2,0)$ under $U(1)_g \times U(1)_R$ with the D$_c$ boundary condition
\item 1 chiral multiplet charged $(0,0)$ under $U(1)_g \times U(1)_R$ with the D boundary condition
\item UV CS level for $U(1)_g$, $k_{gg}=\frac{3}{2}$
\item UV mixed CS level for $U(1)_g$ and $U(1)_R$, $k_{gR}=-\frac{3}{2}$.
\end{itemize}

As another example, the homological block of Brieskorn sphere $\Sigma(2,3,7)$ is given by 
\begin{align}
Z^{SL(2,\mathbb{C})}_{\Sigma(2,3,7)} \simeq \widetilde{\Psi}_{42}^{(1)}(q) + \widetilde{\Psi}_{42}^{(41)}(q) + \widetilde{\Psi}_{42}^{(55)}(q) + \widetilde{\Psi}_{42}^{(71)}(q)
\label{hb237}
\end{align}
and it can also be expressed as \cite{CCFGH}
\begin{align}
q^{\frac{1}{168}} \sum_{k=0}^{\infty} \frac{(-1)^{k} q^{\frac{k^2}{2}+\frac{k}{2}}}{(q^{k+1};q)_k}	\,	.
\label{hb2371}
\end{align}
The corresponding 3d $\mathcal{N}=2$ theory is
\begin{itemize}[leftmargin=5mm]
\item $U(1)_g$ vector multiplet with the $\mathcal{D}$ boundary condition
\item 1 chiral multiplet charged $(1,2)$ under $U(1)_g \times U(1)_R$ with the N boundary condition
\item 1 chiral multiplet charged $(-2,0)$ under $U(1)_g \times U(1)_R$ with the D$_c$ boundary condition
\item 1 chiral multiplet charged $(0,0)$ under $U(1)_g \times U(1)_R$ with the D boundary condition
\item UV CS level for $U(1)_g$, $k_{gg}=-\frac{1}{2}$
\item UV mixed CS level for $U(1)_g$ and $U(1)_R$, $k_{gR}=\frac{1}{2}$.
\end{itemize}

Though the homological block for closed 3-manifolds is expressed in general as a sum of characters, we expect that as a whole it can also be expressed as a version of the quiver form as in the case of $\Sigma(2,3,5)$ and $\Sigma(2,3,7)$, which admit a direct interpretation in terms of 3d $\mathcal{N}=2$ theory as discussed above.

%%%%%%%%%%%%%%%%%%%%%%%%%%%%%%%%%%%%%%%%%%%%%%%%%%%%%%%%%%%%%%%%%%%%%%%%%%%%%%%%%%%%

\subsubsection{Orientation reversal of 3-manifold}

We consider orientation reversal $-M_3$ of the 3-manifold $M_3$ and the corresponding operation in the 3d $\mathcal{N}=2$ theory $T[M_3]$ with boundary conditions.
It is known that 
\begin{align}
Z_{CS}(M_3,k) = Z_{CS}(-M_3, -k)	\,	,
\end{align}
which can be seen formally in the path integral and has also been discussed at the perturbative level \cite{Dimofte-Gukov-Lenells-Zagier}.
Therefore, by replacing the Chern-Simons level $k$ with $-k$, or $q \rightarrow q^{-1}$ in the partition function for $M_3$ and re-expanding the resulting expression as $q$-series, we can at least formally have the partition function for the orientation-reversed 3-manifold $-M_3$.
This is also the case for homological blocks \cite{CCFGH, Chung-seifert}.	\\

The orientation reversal of $M_3$ or $q \leftrightarrow q^{-1}$ is used for the calculation of the index of $T[M_3]$ on $S^2 \times_q S^1$.
It is expected that the index of the 3d $\mathcal{N}=2$ theory $T[M_3]$ on $S^2 \times_q S^1$ is given by
\begin{align}
\mathcal{I}(q) = \sum_{b} |\mathcal{W}_b| \widehat{Z}_b(q) \widehat{Z}_b(q^{-1})
\label{indfact}
\end{align}
where $b$ denotes abelian flat connections and $|\mathcal{W}_b|$ is the order of stabilizer subgroup of the Weyl group for $b$ \cite{Gukov-Pei-Putrov-Vafa}.
We see that $\widehat{Z}_b(q^{-1})$ is used in the calculation of the index \eqref{indfact}.

In addition, $q \rightarrow q^{-1}$ takes the false theta function, which for example appears in the homological blocks for Seifert manifolds, on the upper half plane $\mathbb{H}$ of $\tau$ of $q=e^{\frac{2\pi i}{k}} = e^{2\pi i \tau}$ or inside the unit circle $|q|<1$ to the mock theta function on the lower half plane $\mathbb{H}^{-}$ of $\tau$ or outside the unit circle $|q|>1$, vice versa \cite{CCFGH}.	\\

Since the operation $q \leftrightarrow q^{-1}$ and re-expansion as $q$-series can be done for certain $q$-hypergeometric series, we may ask what the operation would be at the level of 3d $\mathcal{N}=2$ theories $T[M_3]$ with boundary conditions.\footnote{In general, there is ambiguity to determine the $q$-hypergeometric series outside unit circle $|q|>1$ from inside unit circle $|q|<1$, and vice versa \cite{CCFGH}.}
Here, we consider two examples $\Sigma(2,3,5)$ and $\Sigma(2,3,7)$ discussed above where both $q$-series inside and outside of the unit circle are known \cite{CCFGH}.

The homological block \eqref{hbs2351} for $\Sigma(2,3,5)$ is convergent both for $|q|<1$ and $|q|>1$.
By taking $q \rightarrow q^{-1}$ in \eqref{hbs2351} and re-expanding as $q$-series,
\begin{align}
q^{-\frac{1}{120}} \bigg( 2 - \sum_{k=0}^{\infty} \frac{q^{k}}{(q^{k+1};q)_{k}} \bigg) = q^{-\frac{1}{120}}\big( 2- \chi_0(q) \big)
\label{hb2352}
\end{align}
is obtained where $\chi_0$ is the order 5 mock theta function.
The 3d $\mathcal{N}=2$ theory whose half-index is $\chi_0$ is
\begin{itemize}[leftmargin=5mm]
\item $U(1)_g$ vector multiplet with the $\mathcal{D}$ boundary condition
\item 1 chiral multiplet charged $(1,2)$ under $U(1)_g \times U(1)_R$ with the N boundary condition
\item 1 chiral multiplet charged $(-2,0)$ under $U(1)_g \times U(1)_R$ with the D$_c$ boundary condition
\item 1 chiral multiplet charged $(0,0)$ under $U(1)_g \times U(1)_R$ with the D boundary condition
\item UV CS level for $U(1)_g$, $k_{gg}=-\frac{3}{2}$
\item UV mixed CS level for $U(1)_g$ and $U(1)_R$, $k_{gR}=\frac{3}{2}$
\end{itemize}

We can also consider the case of $\Sigma(2,3,7)$.
The homological block \eqref{hb2371} for $\Sigma(2,3,7)$ is also convergent both for $|q|<1$ and $|q|>1$.
After taking $q \rightarrow q^{-1}$, \eqref{hb2371} becomes
\begin{align}
q^{-\frac{1}{168}} \sum_{k=0}^{\infty} \frac{q^{k^2}}{(q^{k+1};q)_k} = q^{-\frac{1}{168}} F_0(q) 
\label{hb2372}
\end{align}
where $F_0$ is the order 7 mock theta function.
The corresponding 3d $\mathcal{N}=2$ theory is
\begin{itemize}[leftmargin=5mm]
\item $U(1)_g$ vector multiplet with the $\mathcal{D}$ boundary condition
\item 1 chiral multiplet charged $(1,2)$ under $U(1)_g \times U(1)_R$ with the N boundary condition
\item 1 chiral multiplet charged $(-2,0)$ under $U(1)_g \times U(1)_R$ with the D$_c$ boundary condition
\item 1 chiral multiplet charged $(0,0)$ under $U(1)_g \times U(1)_R$ with the D boundary condition
\item UV CS level for $U(1)_g$, $k_{gg}=\frac{1}{2}$
\item UV mixed CS level for $U(1)_g$ and $U(1)_R$, $k_{gR}=-\frac{1}{2}$
\end{itemize}

For the examples above,\footnote{Around the completion of this work, we noticed that the examples, $\chi_0(q)$ and $F_0(q)$, were discussed in the context of 3d $\mathcal{N}=2$ theories in \cite{Jockers:2021omw}, but the boundary conditions discussed here that involves the D$_c$ boundary conditions are different from the boundary conditions in \cite{Jockers:2021omw}, which contains certain types of Wilson lines or limits.} we see that upon the orientation reversal of $M_3$ the UV (mixed) Chern-Simons levels of the 3d $\mathcal{N}=2$ theory change sign while other data stay the same.	\\

This behavior of the 3d $\mathcal{N}=2$ theory upon $q \rightarrow q^{-1}$ can be seen in a general setup. 
We first consider a quiver form in the standard expression \eqref{gendtn} with $x_j=1$, 
\begin{align}
\mathcal{P}_Q(\mathbf{x}=1,q) = \sum_{m_1, \ldots m_L \geq 0} q^{\frac{1}{2} \sum_{i,j} C_{ij} m_i m_j} (-q^{\frac{1}{2}})^{\sum_{j=1}^{L}p_j m_j} \frac{1}{\prod_{j=1}^{L}(q;q)_{m_j}}
\label{gendtnx1}
\end{align}
and assume that taking $q \rightarrow q^{-1}$ on \eqref{gendtnx1} and re-expansion as $q$-series also gives a convergent series.

The 3d $\mathcal{N}=2$ theory whose half-index is \eqref{gendtnx1} is given by the theory obtained from \eqref{gendtn} with $U(1)_x$ turned off.
In this case, the UV CS levels can be extracted from 
\begin{align}
\sum_{i,j=1}^{L} C_{ij} \mathbf{f}_i \mathbf{f}_j  + 2\mathbf{f}_R \sum_{j=1}^{L} p_j \mathbf{f}_j - \bigg(\sum_{j=1}^{L} \frac{1}{2}\mathbf{f}_j^2 + \mathbf{f}_j \mathbf{f}_R\bigg)
\label{cslevelq}
\end{align}
where the contribution of $(q;q)_{m_j}^{-1}$ to the anomaly polynomial is $\frac{1}{2}\mathbf{f}_j^2 + \mathbf{f}_j \mathbf{f}_R$.

Meanwhile, taking $q \rightarrow q^{-1}$ in $(q;q)_m^{-1}$, we obtain 
\begin{align}
\frac{1}{(q^{-1};q^{-1})_m} = q^{\frac{1}{2}m^2} (-q^{\frac{1}{2}})^m \frac{1}{(q;q)_m}
\label{revq}
\end{align}
The contribution of \eqref{revq} to the anomaly polynomial is $-\left(\frac{1}{2} \mathbf{f}^2 + \mathbf{f} \, \mathbf{f}_R\right)$.
In addition, we also have $q^{-\frac{1}{2} \sum_{i,j} C_{ij} m_i m_j} (-q^{\frac{1}{2}})^{-\sum_{j=1}^{L}p_j m_j}$.
So the resulting expression upon $q \rightarrow q^{-1}$ is
\begin{align}
\sum_{m_1, \ldots m_L \geq 0} q^{-\frac{1}{2} \sum_{i,j} C_{ij} m_i m_j} (-q^{\frac{1}{2}})^{-\sum_{j=1}^{L}p_j m_j} \prod_{j=1}^{L} \frac{q^{\frac{1}{2}m_j^2} (-q^{\frac{1}{2}})^{m_j}}{(q;q)_{m_j}}	\,	.
\label{gendtnx1invq}
\end{align}
Therefore, the UV CS levels of the 3d $\mathcal{N}=2$ theory whose half-index is \eqref{gendtnx1invq} are extracted from the anomaly polynomial 
\begin{align}
-\sum_{i,j=1}^{L} C_{ij} \mathbf{f}_i \mathbf{f}_j  - 2\mathbf{f}_R \sum_{j=1}^{L} p_j \mathbf{f}_j + \bigg(\sum_{j=1}^{L} \frac{1}{2}\mathbf{f}_j^2 + \mathbf{f}_j \mathbf{f}_R\bigg)	\,	,
\end{align}
and this has an opposite sign of \eqref{cslevelq}.
Therefore, upon $q\rightarrow q^{-1}$ in \eqref{gendtnx1}, other contents of the 3d $\mathcal{N}=2$ theory stay the same but the UV (mixed) CS levels of the 3d $\mathcal{N}=2$ theory change the signs.
For the case that there are additional $q$-Pochhammer symbols $(q;q)_n$ with $n < \infty$ in the numerator or the denominator of \eqref{gendtnx1}, as in the case of $\Sigma(2,3,5)$ and $\Sigma(2,3,7)$, we can also see that only the UV (mixed) CS levels of the 3d $\mathcal{N}=2$ theory change the signs upon $q \rightarrow q^{-1}$ while other data of the theory stay the same.

It would be interesting to study the relation of the theories $T[M_3]$ or $T[M_3\backslash K]$ upon $q \leftrightarrow q^{-1}$ and possible applications on the study of quantum modular forms.

%%%%%%%%%%%%%%%%%%%%%%%%%%%%%%%%%%%%%%%%%%%%%%%%%%%%%%%%%%%%%%%%%%%%%%%%%%%%%%%%%%%%
%%%%%%%%%%%%%%%%%%%%%%%%%%%%%%%%%%%%%%%%%%%%%%%%%%%%%%%%%%%%%%%%%%%%%%%%%%%%%%%%%%%%
%%%%%%%%%%%%%%%%%%%%%%%%%%%%%%%%%%%%%%%%%%%%%%%%%%%%%%%%%%%%%%%%%%%%%%%%%%%%%%%%%%%%

\subsection{3d-3d correspondence for Jones polynomial}
\label{ssec:jones}

The homological block $\mathcal{F}_K(x,q)$ for a knot complement $S^3 \backslash K$ can be regarded as analytic continuation of the colored Jones polynomial where both the level $k$ of $q=e^{\frac{2\pi i }{k}}$ and the color $r$ of $x=q^r$ are analytically continued \cite{Gukov-Manolescu, Park-largecolor, Chung-resurg}.
If taking $r$ of $x=q^r$ to be a non-negative integer while $q$ be still analytically continued in $\mathcal{F}_K(x,q)$, the colored Jones polynomial $J^{\mathcal{S}^r}_K(q)$ for $r=n-1$ dimensional representation with analytically continued $q$ is obtained.
If we can obtain a 3d $\mathcal{N}=2$ theory from the balanced expansion $\mathcal{F}_K(x,q)$ of homological blocks in general, then by considering the case of $x=q^{n-1}$ we would be able to obtain the 3d $\mathcal{N}=2$ theory whose half-index is the colored Jones polynomial.
However, the standard quiver form \eqref{gendt} or \eqref{gendtn} that we extract the 3d $\mathcal{N}=2$ theory is not obvious for homological blocks in the balanced expansion.\footnote{For the torus knot, say $T^{l}(2,2p+1)$, the positive expansion with $x=q^{n-1}$ gives the colored Jones polynomial, so in this case, we can directly work on the positive expansion in the quiver form with $x=q^{n-1}$ and can obtain the 3d $\mathcal{N}=2$ theory.}
Meanwhile, the expression of the colored Jones polynomial as a quiver form with additional $q$-Pochhammer symbols are available for some infinite families of knots.
So from such expressions, we can obtain a 3d $\mathcal{N}=2$ theory whose half-index is the colored Jones polynomial $J_K(n,q)$ for the $n$-dimensional representation of $G=SU(2)$.	\\

For example, the colored Jones polynomial for the twist knot $T_{K_{p \gtrless 0}}$ and the left-handed torus knot $T^{l}(2,2p+1)$ admit a quiver form with some additional $q$-Pochhammer symbols, which are expressed in the reduced normalization as \cite{Masbaum2003,FGSS-AD,Hikami-hecke,Hikami-torusknot}
\begin{align}
J_{K_{p >0}}(n,q) &= \sum^{\infty}_{0 \leq s_1 \leq \ldots \leq s_p} q^{s_p} (q^{1-n};q)_{s_p} (q^{1+n};q)_{s_p} \prod_{i=1}^{p-1} q^{s_i(s_i+1)} \frac{(q;q)_{s_{i+1}}}{(q;q)_{s_{i}}(q;q)_{s_{i+1}-s_{i}}}	
\label{jtwp}\\
J_{K_{-p <0}}(n,q) &= \sum^{\infty}_{0 \leq s_1 \leq \ldots \leq s_p} (-1)^{s_p} q^{-\frac{1}{2}s_p(s_p+1)} (q^{1-n};q)_{s_p} (q^{1+n};q)_{s_p} \prod_{i=1}^{p-1} q^{s_i(s_i+1)} \frac{(q;q)_{s_{i+1}}}{(q;q)_{s_{i}}(q;q)_{s_{i+1}-s_{i}}}
\label{jtwm}	\\
J_{T^{l}(2,2p+1)}(n,q) &= q^{p(n-1)} \sum^{\infty}_{0 \leq s_1 \leq \ldots \leq s_p} q^{-\frac{1}{2}s_p^2 } (-q^{\frac{1}{2}})^{s_p} q^{(2n-1)\sum^{p}_{i=1} s_i} (q^{1-n};q)_{s_p}  
\prod^{p-1}_{i=1} q^{-s_i s_{i+1}}\frac{(q;q)_{s_{i+1}}}{(q;q)_{s_{i}}(q;q)_{s_{i+1}-s_{i}}}
\label{jtorus}
\end{align}
for the $n$-dimensional representation of $G=SU(2)$.
The sum terminates at $s_p =n-1$ in \eqref{jtwp}, \eqref{jtwm}, and \eqref{jtorus}.
In terms of $m_j = s_j - s_{j-1}$, $j=1, \ldots, p$ with $s_0=0$, they can be expressed as
\begin{align}
\begin{split}
J_{K_{p >0}}(n,q) &= \sum^{\infty}_{m_1, \ldots, m_p \geq 0} 
q^{\sum_{j=1}^{p}(p-j)m_j^2 + 2 \sum_{i=1}^{p-1} \sum_{j=i+1}^{p}(p-j) m_i m_j} (-q^{\frac{1}{2}})^{2\sum_{j=1}^{p} (p-j+1) m_j}	\\
&\hspace{25mm}\times \frac{(q^{1-n};q)_{\infty}}{(q^{1-n+\sum_{j=1}^{p}m_j};q)_{\infty}} \frac{(q^{1+n};q)_{\infty}}{(q^{{1+n}+\sum_{j=1}^{p}m_j};q)_{\infty}} \frac{(q;q)_{\sum_{j=1}^{p}m_j}}{\prod_{i=1}^{p}(q;q)_{m_j}}	
\label{jtwp2}
\end{split}	\\
\begin{split}
J_{K_{-p <0}}(n,q) &= 
\sum^{\infty}_{m_1, \ldots, m_p \geq 0} q^{\frac{1}{2}\sum_{j=1}^{p} (2p-2j-1)m_j^2 + \sum_{i=1}^{p-1} \sum_{j=i+1}^{p} (2p-2j-1) m_i m_j} (-q^{\frac{1}{2}})^{\sum_{j=1}^{p}(2p-2j-1)m_j}	\\
&\hspace{25mm} \times \frac{(q^{1-n};q)_{\infty}}{(q^{1-n+\sum_{j=1}^{p}m_j};q)_{\infty}} \frac{(q^{1+n};q)_{\infty}}{(q^{{1+n}+\sum_{j=1}^{p}m_j};q)_{\infty}} \frac{(q;q)_{\sum_{j=1}^{p}m_j}}{\prod_{i=1}^{p}(q;q)_{m_j}}	
\label{jtwm2}
\end{split}	\\
\begin{split}
J_{T^{l}(2,2p+1)}(n,q) &= q^{p(n-1)} 
\sum^{\infty}_{m_1, \ldots, m_p \geq 0} q^{\frac{1}{2} \sum_{j=1}^{p} (-2p+2j-1)m_j^2 -2 \sum_{i=1}^{p-1} \sum_{j=i+1}(p-j+1)m_i m_j} (-q^{\frac{1}{2}})^{\sum_{j=1}^{p} (-2p+2j-1)m_j} 	\\
&\hspace{25mm} \times q^{2n \sum_{j=1}^{p} (p-j+1)m_j} \frac{(q^{1-n};q)_{\infty}}{(q^{1-n+\sum_{j=1}^{p}m_j};q)_{\infty}} \frac{(q;q)_{\sum_{j=1}^{p}m_j}}{\prod_{j=1}^{p}(q;q)_{m_j}}	\,	.
\label{jtorus2}
\end{split}
\end{align}
When $p \geq 2$, there is no cancellation of the last factor $\frac{(q;q)_{\sum_{j=1}^{p}m_j}}{\prod_{i=1}^{p}(q;q)_{m_j}}$, and we have a quiver form with additional $q$-Pochhammer symbols.
As in the case of closed 3-manifolds in section \ref{ssec:lcft}, the form of \eqref{jtwp2}, \eqref{jtwm2}, and \eqref{jtorus2} are not exactly the quiver form of \eqref{gendt}, but we can still obtain 3d $\mathcal{N}=2$ theories whose half-indices are them.

Regarding the factor $q^{\pm n}$ in \eqref{jtwp2}, \eqref{jtwm2}, and \eqref{jtorus2}, one way to interpret it would be to consider a vortex of charge $n$ for a global symmetry \cite{Dimofte-Gaiotto-Paquette}, which we denote by $U(1)_v$.
For example, we can interpret the factor $(q^{1-n};q)_\infty$ as the half-index of the 3d chiral multiplet in the D$_c$ boundary condition, which is charged $(-1,0)$ under $U(1)_v \times U(1)_R$ symmetry, with a vortex line with the vortex charge $n$.\footnote{We note that we deliberately choose the expression for the $n$-dimensional representation, which gives a form $(q^{1-n};q)_\infty$ instead of $(q^{-r};q)_\infty$ where $r=n-1$.
This is because we would like to have R-charge 0 for the chiral multiplet with the D$_c$ boundary condition, which is necessary in order to preserve the $U(1)_R$ symmetry. }	\\

For a twist knot $K_{p>0}$ with $p>1$, the 3d $\mathcal{N}=2$ theory $T[J_{K_{p>0}}(n,q)]$ is given by
\begin{itemize}[leftmargin=5mm]
\item $U(1)_j$, $j =1, \ldots, p$ vector multiplets with the $\mathcal{D}$ boundary condition
\item $p$ chiral multiplet $\Phi_i$, $i=1, \ldots, p$ charged $(-\delta_{ij},0,0)$ under $U(1)_j\text{'s} \times U(1)_v \times U(1)_R$ with the D$_c$ boundary condition
\item 2 chiral multiplet charged $(0,1,0)$, and $(0,-1,0)$ under $U(1)^p \times U(1)_v \times U(1)_R$ with the D$_c$ boundary condition
\item 1 chiral multiplet charged $(0,0,0)$ under $U(1)^p \times U(1)_v \times U(1)_R$ with the D boundary condition
\item 3 chiral multiplet charged $(1,0,2)$, $(1,1,2)$, and $(1,-1,2)$ under $U(1)^p \times U(1)_v \times U(1)_R$ with the N boundary condition
\item UV (mixed) CS levels for $U(1)^p$ $-$ $U(1)^p$ :
$
\begin{pmatrix}
2p-1			&2p-\frac{5}{2}	&2p-\frac{9}{2}	&\ldots	&\frac{7}{2}	&\frac{3}{2}	\\
2p-\frac{5}{2}	&2p-3		&2p-\frac{9}{2}	&\ldots	&\frac{7}{2}	&\frac{3}{2}	\\
2p-\frac{9}{2}	&2p-\frac{9}{2}	&2p-5		&\ldots	&\frac{7}{2}	&\frac{3}{2}	\\
\vdots		&\vdots		&\vdots		&\ddots	&			&	\\
\frac{7}{2}		&\frac{7}{2}	&\frac{7}{2}	&		&3			&\frac{3}{2}	\\
\frac{3}{2}		&\frac{3}{2}	&\frac{3}{2}	&		&\frac{3}{2}	&1	
\end{pmatrix}
$
\item UV mixed CS levels for $U(1)^p$ $-$ $U(1)_R$ : $(2p+1, 2p-1, \ldots, 3)$
\item vortex line for $U(1)_v$ symmetry with the vortex charge $n$.
\end{itemize}
The 3d $\mathcal{N}=2$ theories $T[J_{K}(n,q)]$ for $K_{p=1}$ and for the other twist knot $K_{-p<0}$ with $-p\leq-1$ and torus knot $T^{l}(2,2p+1)$ with $p\geq1$ are available in Appendix \ref{app:jones}.
We can also obtain a 3d $\mathcal{N}=2$ theory for the Jones polynomial of all double-twist knots in \cite{Lauridsen, Lovejoy-Osburn2021,Lovejoy-Osburn2019}.
As discussed in the case for infinite families of knots here and in Appendix \ref{app:jones}, it is expected for a general knot $K$ that we can obtain the 3d $\mathcal{N}=2$ theory $T[J_{K}(n,q)]$ with the 2d $\mathcal{N}=(0,2)$ boundary conditions and the flavor vortex line whose half-index is the colored Jones polynomial $J_{K}(n,q)$ with analytically continued $q$.	\\

We can also obtain 3d $\mathcal{N}=2$ theory $T[P^{SU(N)}_K(n,q)]$ or $T[P^{SU(N)}_K(n,q,t)]$ for the case $G=SU(N)$ whose half-index is the specialization $a=q^N$ of the $\mathcal{S}^{n-1}$ colored HOMFLY polynomial or the $\mathcal{S}^{n-1}$ colored superpolynomial.
For example, the quiver form with additional $q$-Pochhammer symbols are available for $\mathcal{S}^{n-1}$-colored superpolynomials for the $(2,2p+1)$ torus knot \cite{FGS-VC}, the $(2,2p)$ torus knot \cite{GNSSS}, twist knot \cite{FGSS-AD, NRZS}, and from them we can obtain the 3d $\mathcal{N}=2$ theory whose half-index is the $\mathcal{S}^{n-1}$-colored HOMFLY polynomial or superpolynomial at $a=q^N$.	\\

In the context of the M-theory configuration, such $T[J_{K}(n,q)]$ would be realized by adding additional M2 branes.
A brane construction that gives the analytically continued Jones polynomial with a given representation $\mathcal{R}$ is given by \cite{GKRY, GNSSS, Gukov-Pei-Putrov-Vafa}
\begin{align}
\begin{tabular}{r c c c c c c}
\text{space-time}		&	&	$\mathbb{R}$ 	&$\times$ 	&$TN$ 	&$\times$ 	&$T^* M_3$	\\
\text{M5 branes}		&	&	$\mathbb{R}$ 	&$\times$ 	&$D^2$ 	&$\times$ 	&$M_3$	\\
\text{M2 branes}		&	&	$\mathbb{R}$ 	&$\times$ 	&$pt$ 	&$\times$ 	&$T^*K$	\\
\end{tabular}
\label{m2conf}
\end{align}
where $pt$ is a point at the center of $D^2$.
The representation is specified by the way that M2 branes end on $N$ M5 branes. 
From such configuration, partitions of the number of M2 branes by $N$ give the Young tableaux for $G=U(N)$, which gives the irreducible representation of $U(N)$.
If M2 branes are equally distributed on $N$ M5 branes, the column of $N$ boxes can be taken out from the Young tableaux, and the Young tableaux of $G=SU(N)$ is obtained, which gives the representation $\mathcal{R}$.
There is another way to realize a knot with a given representation by adding additional M5 branes supported on $\mathbb{R} \times L_K$ and the cotangent space to $D^2$ at the point $pt$.
Both of them lead to a Wilson loop in Chern-Simons theory, and $T[J_K(n,q)]$ discussed above would be one of 3d $\mathcal{N}=2$ theories whose half-index is the colored Jones polynomial.

%%%%%%%%%%%%%%%%%%%%%%%%%%%%%%%%%%%%%%%%%%%%%%%%%%%%%%%%%%%%%%%%%%%%%%%%%%%%%%%%%%%%
%%%%%%%%%%%%%%%%%%%%%%%%%%%%%%%%%%%%%%%%%%%%%%%%%%%%%%%%%%%%%%%%%%%%%%%%%%%%%%%%%%%%
%%%%%%%%%%%%%%%%%%%%%%%%%%%%%%%%%%%%%%%%%%%%%%%%%%%%%%%%%%%%%%%%%%%%%%%%%%%%%%%%%%%%
%%%%%%%%%%%%%%%%%%%%%%%%%%%%%%%%%%%%%%%%%%%%%%%%%%%%%%%%%%%%%%%%%%%%%%%%%%%%%%%%%%%%
%%%%%%%%%%%%%%%%%%%%%%%%%%%%%%%%%%%%%%%%%%%%%%%%%%%%%%%%%%%%%%%%%%%%%%%%%%%%%%%%%%%%

\acknowledgments{I would like to thank Sergei Gukov for comments on the draft.
I also would like to thank the Korea Institute for Advanced Study (KIAS) for hospitality at the final stage of this work.
This research was supported by the 2023 scientific promotion program funded by Jeju National University.}

%%%%%%%%%%%%%%%%%%%%%%%%%%%%%%%%%%%%%%%%%%%%%%%%%%%%%%%%%%%%%%%%%%%%%%%%%%%%%%%%%%%%
%%%%%%%%%%%%%%%%%%%%%%%%%%%%%%%%%%%%%%%%%%%%%%%%%%%%%%%%%%%%%%%%%%%%%%%%%%%%%%%%%%%%
%%%%%%%%%%%%%%%%%%%%%%%%%%%%%%%%%%%%%%%%%%%%%%%%%%%%%%%%%%%%%%%%%%%%%%%%%%%%%%%%%%%%
%%%%%%%%%%%%%%%%%%%%%%%%%%%%%%%%%%%%%%%%%%%%%%%%%%%%%%%%%%%%%%%%%%%%%%%%%%%%%%%%%%%%
%%%%%%%%%%%%%%%%%%%%%%%%%%%%%%%%%%%%%%%%%%%%%%%%%%%%%%%%%%%%%%%%%%%%%%%%%%%%%%%%%%%%

\begin{appendices}

%%%%%%%%%%%%%%%%%%%%%%%%%%%%%%%%%%%%%%%%%%%%%%%%%%%%%%%%%%%%%%%%%%%%%%%%%%%%%%%%%%%%
%%%%%%%%%%%%%%%%%%%%%%%%%%%%%%%%%%%%%%%%%%%%%%%%%%%%%%%%%%%%%%%%%%%%%%%%%%%%%%%%%%%%
%%%%%%%%%%%%%%%%%%%%%%%%%%%%%%%%%%%%%%%%%%%%%%%%%%%%%%%%%%%%%%%%%%%%%%%%%%%%%%%%%%%%

\section{Quiver forms and refinement}
\label{app:ref}

The refined knot polynomial or the Poincar\'e polynomial has additional parameter $t$ and physically this parameter is associated to $U(1)_R$ symmetry.
So we also denote $U(1)_R$ by $U(1)_t$.
The knot polynomial that is not refined is expressed as
\begin{align}
P_K(q) = \text{Tr}_{\mathcal{H}} \, (-1)^j q^i
\label{unrk}
\end{align}
where $\mathcal{H}$ is a knot homology or the space of BPS states (\textit{c.f.} \eqref{grhind}).
A refined version of the knot polynomial is expressed as
\begin{align}
P_K(q,t) = \text{Tr}_{\mathcal{H}} \, t^j q^i	\,	.
\label{pind}
\end{align}
and by taking $t=-1$ the knot polynomial \eqref{unrk} is obtained.

The parameter $t$ cannot be turned on in general in the supersymmetric index \cite{Witten-M5knots}.
Meanwhile, when $M_3$ is a Seifert manifold which admits a semi-free $U(1)$ action on $M_3$, there is a nowhere vanishing vector field, and there is $U(1)_S$ symmetry which is the rotation of the two-plane that is orthogonal to the vector field in the fiber of $T^* M_3$.
Then, the $U(1)_\beta$ symmetry
\begin{align}
U(1)_\beta = U(1)_S - U(1)_t
\end{align}
is a global symmetry and the parameter $\beta$ can be turned on in the index \cite{AS-refinedCS}
\begin{align}
\mathcal{I}(q,t) = \text{Tr}_{\mathcal{H}} \, (-1)^R q^{\frac{R}{2}+J} \beta^{R_{S}-R}	\,	.
\label{betaind}
\end{align}
The supersymmetric index \eqref{betaind} gives the Poincar\'e polynomial \eqref{pind} for the Seifert knot when $\beta=-t^{-1}$ if the BPS states that contribute to \eqref{betaind} have the charge $R_S=0$ and there is no cancellation in \eqref{betaind} due to $(-1)^R$, which are expected to hold \cite{AS-refinedCS, Gukov-Putrov-Vafa}.	\\

As an example, we consider the left-handed torus knots $T^{l}(2,2p+1)$, which is a Seifert knot.
The $\mathcal{S}^r$-colored superpolynomial of the torus knot $T^{l}(2,2p+1)$ is given by \cite{FGSS-AD}
\begin{align}
\begin{split}
P^{\mathcal{S}^r}_{T^{l}(2,2p+1)}(a,q,t) &= a^{pr} q^{-pr} \sum_{0 \leq k_p \leq \ldots \leq k_1}^{\infty} q^{-\sum_{i=1}^{p}k_{i-1} k_{i}}q^{(2r+1) \sum_{j=1}^p k_j} (-t)^{2 \sum_{j=1}^p k_j} 
(aq^{-1}(-t);q)_{k_1}	\\
&\hspace{35mm}\times \prod_{j=0}^{p-1}
\frac{(q;q)_{k_j}}{(q;q)_{k_{j+1}} (q;q)_{k_j-k_{j+1}}}
\label{splt}
\end{split}
\end{align}
where $k_0=r$ and this sum terminates at $k_1=r$.

By taking $x=q^r$, up to the prefactor $a^{pr} q^{-pr}$, the $a$, $-t$-deformed homological block for the torus knot can be expressed as \cite{EGGKPS}
\begin{align}
\begin{split}
F_{T^{l}(2,2p+1)}(x,a,q,t) =& \sum^{\infty}_{0 \leq k_1 \leq \ldots \leq k_p} q^{-\frac{1}{2}k_p^2 - \sum_{j=2}^{p} k_{j-1} k_{j} } (-q^{\frac{1}{2}})^{k_p+2\sum_{j=1}^p k_j} x^{2 \sum_{j=1}^p k_j} (-t)^{2 \sum_{j=1}^p k_j} 	\\
&\times (x^{-1};q)_{k_{p}} (aq^{-1}(-t);q)_{k_{p}} \frac{(q;q)_{k_{p}}}{(q;q)_{k_{p}-k_{p-1}} (q;q)_{k_{p-1}-k_{p-2}} \cdots (q;q)_{k_{2}-k_{1}} (q;q)_{k_1}}	\,	.
\end{split}
\end{align}
With $m_j=k_j-k_{j-1}$, $j=1, \ldots, p$, we have
\begin{align}
\begin{split}
F_{T^{l}(2,2p+1)}(x,a,q,t) =& \sum^{\infty}_{m_1, \ldots, m_p \geq 0} 
q^{\frac{1}{2} \sum_{j=1}^{p} (-2p+2j-1)) m_j^2  -2 \sum_{i=1}^{p-1} \sum_{j=i+1}^{p} (p-j+1)m_i m_j} 	\\
&\hspace{10mm} \times x^{2\sum_{j=1}^{p}(p-j+1)m_j} (-t)^{2\sum_{j=1}^{p}(p-j+1)m_j} (-q^{\frac{1}{2}})^{\sum_{j=1}^{p} (2p-2j+3)m_j } 	\\
&\hspace{10mm} \times (x^{-1};q)_{\sum_{i=1}^{p}m_i} (aq^{-1}(-t);q)_{\sum_{i=1}^{p}m_i} \frac{(q;q)_{\sum_{i=1}^{p}m_i}}{ \prod_{i=1}^{p} (q;q)_{m_i} }	\,	.
\end{split}
\label{lfktorus}
\end{align}

The 3d $\mathcal{N}=2$ theory for the $a$, $-t$-deformed homological block \eqref{lfktorus} is given by 
\begin{itemize}[leftmargin=5mm]
\item $U(1)_j$, $j=1, \ldots, p$ vector multiplets with the $\mathcal{D}$ boundary condition
\item $p$ chiral multiplets $\Phi_i$, $i=1, \ldots, p$ charged $(-\delta_{ij},0,0,0,0)$ under $U(1)_j\text{'s} \times U(1)_x \times U(1)_a \times U(1)_t \times U(1)_R$ with the D$_{c}$ boundary condition
\item 3 chiral multiplets charged $(0,0,0,0,0)$, $(0,0,-1,-1,4)$, $(0,1,0,0,2)$ under $U(1)^p \times U(1)_x \times U(1)_a \times U(1)_t \times U(1)_R$ with the D boundary condition
\item 3 chiral multiplets charged $(1,0,0,0,2)$, $(1,0,1,1,-2)$, $(1,-1,0,0,0)$ under $U(1)^p \times U(1)_x \times U(1)_a \times U(1)_t \times U(1)_R$ with the N boundary condition
\item UV (mixed) CS level for $U(1)^p$ $-$ $U(1)^p$ : 
$
\begin{pmatrix}
-2(p-1)		&-2p+\frac{7}{2}	&-2p+\frac{11}{2}	&\ldots	&-\frac{5}{2}	&-\frac{1}{2}	\\
-2p+\frac{7}{2}	&-2(p-2)			&-2p+\frac{11}{2}	&\ldots	&-\frac{5}{2}	&-\frac{1}{2}	\\
-2p+\frac{11}{2}	&-2p+\frac{11}{2}	&-2(p-3)			&\ldots	&-\frac{5}{2}	&-\frac{1}{2}	\\
\vdots		&\vdots			&\vdots			&\ddots	&			&	\\
-\frac{5}{2}	&-\frac{5}{2}		&-\frac{5}{2}		&		&-2			&-\frac{1}{2}	\\
-\frac{1}{2}	&-\frac{1}{2}		&-\frac{1}{2}		&		&-\frac{1}{2}	&0	
\end{pmatrix}
$
\item UV mixed CS levels for $U(1)^p$ $-$ $U(1)_x$, $U(1)_a$, $U(1)_t$, and $U(1)_R$
\begin{align*}
\hspace{-5mm}\left(2p-\frac{1}{2},2p-\frac{5}{2}, \ldots, \frac{7}{2},\frac{3}{2}\right)	\,	,	\
\left(2p+\frac{1}{2},2p-\frac{3}{2}, \ldots, \frac{9}{2},\frac{5}{2}\right)	\,	,	\
\left(\frac{1}{2},\frac{1}{2},\ldots, \frac{1}{2},\frac{1}{2}\right)	\,	,	\
\left(2p-1,2p-3, \ldots,3, 1\right).
\end{align*}
\end{itemize}

We worked with the quiver form that contains some additional $q$-Pochhammer symbols.
We can also use the expression for the colored superpolynomial of torus knot $T^{l}(2,2p+1)$ in the form of \eqref{gendtn} obtained in \cite{Kucharski-quiver}, but in this case the size of the quiver is $4p$, which leads to the $U(1)^{4p}$ gauge group.

From \eqref{splt}, we can also obtain the generating function of the colored superpolynomial and the corresponding 3d $\mathcal{N}=2$ theory.

%%%%%%%%%%%%%%%%%%%%%%%%%%%%%%%%%%%%%%%%%%%%%%%%%%%%%%%%%%%%%%%%%%%%%%%%%%%%%%%%%%%%
%%%%%%%%%%%%%%%%%%%%%%%%%%%%%%%%%%%%%%%%%%%%%%%%%%%%%%%%%%%%%%%%%%%%%%%%%%%%%%%%%%%%
%%%%%%%%%%%%%%%%%%%%%%%%%%%%%%%%%%%%%%%%%%%%%%%%%%%%%%%%%%%%%%%%%%%%%%%%%%%%%%%%%%%%

\section{3d $\mathcal{N}=2$ theory for Jones polynomial of twist and torus knots}
\label{app:jones}

Continued from section \ref{ssec:jones}, for the Jones polynomial of the twist knot $K_{-p<0}$ \eqref{jtwm2} 
\begin{align}
\begin{split}
J_{K_{-p <0}}(n,q) &= 
\sum^{\infty}_{m_1, \ldots, m_p \geq 0} q^{\frac{1}{2}\sum_{j=1}^{p} (2p-2j-1)m_j^2 + \sum_{i=1}^{p-1} \sum_{j=i+1}^{p} (2p-2j-1) m_i m_j} (-q^{\frac{1}{2}})^{\sum_{j=1}^{p}(2p-2j-1)m_j}	\\
&\hspace{25mm} \times \frac{(q^{1-n};q)_{\infty}}{(q^{1-n+\sum_{j=1}^{p}m_j};q)_{\infty}} \frac{(q^{1+n};q)_{\infty}}{(q^{{1+n}+\sum_{j=1}^{p}m_j};q)_{\infty}} \frac{(q;q)_{\sum_{j=1}^{p}m_j}}{\prod_{i=1}^{p}(q;q)_{m_j}}
\end{split}
\end{align}
with $p>1$, the 3d $\mathcal{N}=2$ theory $T[J_{K_{-p<0}}(n,q)]$ is given by
\begin{itemize}[leftmargin=5mm]
\item $U(1)_j$, $j =1, \ldots, p$ vector multiplets with the $\mathcal{D}$ boundary condition
\item $p$ chiral multiplets $\Phi_i$, $i=1, \ldots, p$ charged $(-\delta_{ij},0,0)$ under $U(1)_j\text{'s} \times U(1)_v \times U(1)_R$ with the D$_c$ boundary condition
\item 2 chiral multiplets charged $(0,1,0)$ and $(0,-1,0)$ under $U(1)_j\text{'s} \times U(1)_v \times U(1)_R$ with the D$_c$ boundary condition
\item 1 chiral multiplet charged $(0,0,0)$ under $U(1)^p \times U(1)_v \times U(1)_R$ with the D boundary condition
\item 3 chiral multiplets charged $(1,0,2)$, $(1,1,2)$, and $(1,-1,2)$ under $U(1)^p \times U(1)_v \times U(1)_R$ with the N boundary condition
\item UV (mixed) CS levels for $U(1)^p$ $-$ $U(1)^p$'s :
$
\begin{pmatrix}
2p-2			&2p-\frac{7}{2}	&2p-\frac{11}{2}	&\ldots	&\frac{5}{2}	&\frac{1}{2}	\\
2p-\frac{7}{2}	&2p-4		&2p-\frac{11}{2}	&\ldots	&\frac{5}{2}	&\frac{1}{2}	\\
2p-\frac{11}{2}	&2p-\frac{11}{2}	&2p-6		&\ldots	&\frac{5}{2}	&\frac{1}{2}	\\
\vdots		&\vdots		&\vdots		&\ddots	&	&	\\
\frac{5}{2}		&\frac{5}{2}	&\frac{5}{2}			&		&2	&\frac{1}{2}	\\
\frac{1}{2}		&\frac{1}{2}	&\frac{1}{2}			&		&\frac{1}{2}	&0	
\end{pmatrix}
$
\item mixed CS level for $U(1)^p$ $-$ $U(1)_R$ : $\left(2(p-1), 2(p-2), \ldots, 2,0\right)$
\item vortex line for $U(1)_v$ symmetry with the vortex charge $n$.
\end{itemize}

\vspace{3mm}

The expression \eqref{jtorus2} of the Jones polynomial of the left-handed torus knot $T^{l}(2,2p+1)$ 
\begin{align}
\begin{split}
J_{T^{l}(2,2p+1)}(n,q) &= q^{p(n-1)} 
\sum^{\infty}_{m_1, \ldots, m_p \geq 0} q^{\frac{1}{2} \sum_{j=1}^{p} (-2p+2j-1)m_j^2 -2 \sum_{i=1}^{p-1} \sum_{j=i+1}(p-j+1)m_i m_j} (-q^{\frac{1}{2}})^{\sum_{j=1}^{p} (-2p+2j-1)m_j} 	\\
&\hspace{25mm} \times q^{2n \sum_{j=1}^{p} (p-j+1)m_j} \frac{(q^{1-n};q)_{\infty}}{(q^{1-n+\sum_{j=1}^{p}m_j};q)_{\infty}} \frac{(q;q)_{\sum_{j=1}^{p}m_j}}{\prod_{j=1}^{p}(q;q)_{m_j}}
\end{split}
\end{align}
is slightly different from the case of twist knots.
In this case, for example, there is an overall factor $q^{p(n-1)}$.

Since we consider the case that there is a vortex line for $U(1)_v$ with the vortex charge $n$ and then use D$_c$ boundary condition to set $v=1$, $v$ in the half-index is essentially replaced with $q^{-n}$.
Here, we simply use 
\begin{align}
q^{-1}v^{-1} = -\frac{(q^{2}v;q)_\infty (q^{-1}v^{-1};q)_\infty}{(qv;q)_\infty (v^{-1};q)_\infty} = -\frac{\theta((-q^{\frac{1}{2}})^{3} v;q)}{\theta((-q^{\frac{1}{2}}) v;q)}	\,	.
\end{align}
So up to an overall sign we can interpret $q^{n-1}$ as contributions from four chiral multiplets or 2d $\mathcal{N}=(0,2)$ chiral and Fermi multiplet at the boundary.
Since we have $q^{p(n-1)}$, we have $p$ copies of them.
For simplicity, we interpret them as $p$ copies of 2d $\mathcal{N}=(0,2)$ chiral multiplet and Fermi multiplet with charges $(1,2)$ and $(1,3)$ under $U(1)_v \times U(1)_R$, respectively.
Then the 3d $\mathcal{N}=2$ theory $T[J_{T^{l}(2,2p+1)}(n,q)]$ for the torus knot $T^{l}(2,2p+1)$ with $p>1$ is given by
\begin{itemize}[leftmargin=5mm]
\item $U(1)_j$, $j =1, \ldots, p$ vector multiplets with the $\mathcal{D}$ boundary condition
\item $p$ chiral multiplets $\Phi_i$, $i=1, \ldots, p$ charged $(-\delta_{ij},0,0)$ under $U(1)_j\text{'s} \times U(1)_v \times U(1)_R$ with the D$_c$ boundary condition
\item 1 chiral multiplet charged $(0,-1,0)$ under $U(1)^p \times U(1)_v \times U(1)_R$ with the D$_c$ boundary condition
\item 1 chiral multiplet charged $(0,0,0)$ under $U(1)^p \times U(1)_v \times U(1)_R$ with the D boundary condition
\item 2 chiral multiplets charged $(1,0,2)$ and $(1,1,2)$ under $U(1)^p \times U(1)_v \times U(1)_R$ with the N boundary condition
\item $p$ copies of 2d $\mathcal{N}=(0,2)$ chiral and Fermi multiplets with charges $(1,2)$ and $(1,3)$ under $U(1)_v \times U(1)_R$, respectively
\item UV (mixed) CS levels for $U(1)^p$ $-$ $U(1)^p$ : 
$
\begin{pmatrix}
-2p+\frac{3}{2}	&-2p+3			&-2p+5			&\ldots	&-3			&-1	\\
-2p+3		&-2p+\frac{7}{2}	&-2p+5			&\ldots	&-3			&-1	\\
-2p+5		&-2p+5			&-2p+\frac{11}{2}	&\ldots	&-3			&-1	\\
\vdots		&\vdots			&\vdots			&\ddots	&			&	\\
-3			&-3				&-3				&		&-\frac{5}{2}	&-1	\\
-1			&-1				&-1				&		&-1			&-\frac{1}{2}	
\end{pmatrix}
$
\item UV mixed CS levels for $U(1)^p$ $-$ $U(1)_R$ : $\left(-2p+\frac{3}{2}, -2p+\frac{7}{2}, \ldots, -\frac{5}{2},-\frac{1}{2}\right)$
\item UV mixed CS levels for  $U(1)^p$ $-$ $U(1)_v$ : $\left(-2p+\frac{1}{2}, -2p+\frac{5}{2}, \ldots, -\frac{7}{2},-\frac{3}{2}\right)$
\item UV mixed CS level for $U(1)_R$ $-$ $U(1)_v$ : $-2p$
\item UV CS level for $U(1)_R$ : $-8p$
\item vortex line for $U(1)_v$ symmetry with the vortex charge $n$.
\end{itemize}

\vspace{3mm}

So far, we have not included the case that $p=1$.
For completeness, we discuss that it is also possible to obtain the 3d $\mathcal{N}=2$ theories for the case of $p=1$.
When $p=1$, \eqref{jtwp2}, \eqref{jtwm2}, and \eqref{jtorus2} become
\begin{align}
J_{K_1}(n,q) &= \sum_{k \geq 0} q^k (q^{1-n};q)_k (q^{1+n};q)_k		\label{twstkp1}	\\
J_{K_{-1}}(n,q) &= \sum_{k \geq 0} q^{-\frac{1}{2}k^2} (-q^{\frac{1}{2}})^k (q^{1-n};q)_k (q^{1+n};q)_k		\label{twstkm1}	\\
J_{T^{l}(2,3)}(n,q) &= q^{n-1} \sum_{k \geq 0} q^{-\frac{1}{2}k^2} (-q^{\frac{1}{2}})^{-k} q^{2nk} (q^{1-n};q)_k	\,	.	\label{trskl23}
\end{align}
The twist knot $K_1$ is actually the left-handed trefoil knot $T^{l}(2,3)$, and we take the expression \eqref{twstkp1} for it when obtaining the corresponding 3d $\mathcal{N}=2$ theory.

By using $q$-binomial identity, these can be expressed as
\begin{align}
J_{K_1}(n,q) &= \sum_{m_1,m_2 \geq 0} q^{\frac{1}{2}m_1^2} (-q^{\frac{1}{2}})^{3m_1+2m_2} (q^{-n})^{m_1} \frac{(q;q)_{m_1+m_2}(q^{1+n};q)_{m_1+m_2}}{(q;q)_{m_1}(q;q)_{m_2}}	\\
J_{K_{-1}}(n,q) &= \sum_{m_1,m_2 \geq 0} q^{-\frac{1}{2}m_2^2 -m_1m_2} (-q^{\frac{1}{2}})^{2m_1+m_2} (q^{-n})^{m_1} \frac{(q;q)_{m_1+m_2}(q^{1+n};q)_{m_1+m_2}}{(q;q)_{m_1}(q;q)_{m_2}}	\,	.
\end{align}

Thus, the 3d $\mathcal{N}=2$ theory $T[J_{K_1}(n,q)]$ is
\begin{itemize}[leftmargin=5mm]
\item $U(1)_j$, $j =1, 2$ vector multiplets with the $\mathcal{D}$ boundary condition
\item 2 chiral multiplets $\Phi_i$, $i=1,2$ charged $(-\delta_{ij},0,0)$ under $U(1)_j\text{'s} \times U(1)_v \times U(1)_R$ with the $D_c$ boundary condition
\item 1 chiral multiplets charged $(0,1,0)$ under $U(1)^2 \times U(1)_v \times U(1)_R$ with the $D_c$ boundary condition
\item 1 chiral multiplet charged $(0,0,0)$ under $U(1)^2 \times U(1)_v \times U(1)_R$ with the $D$ boundary condition 
\item 2 chiral multiplet charged $(1,0,2)$ and $(1,-1,2)$ under $U(1)^2 \times U(1)_v \times U(1)_R$ with the $N$ boundary condition
\item UV (mixed) CS levels for $U(1)^2$ $-$ $U(1)^2$ : 
$
\begin{pmatrix}
\frac{3}{2}	&1	\\
1		&\frac{1}{2}
\end{pmatrix}
$
\item UV mixed CS levels for $U(1)^2$ $-$ $U(1)_R$ : $\left(\frac{7}{2},\frac{5}{2}\right)$
\item UV mixed CS levels for  $U(1)^2$ $-$ $U(1)_v$ : $\left(\frac{1}{2}, -\frac{1}{2}\right)$
\item vortex line for $U(1)_v$ symmetry with the vortex charge $n$.
\end{itemize}

\vspace{3mm}

The 3d $\mathcal{N}=2$ theory $T[J_{K_{-1}}(n,q)]$ has the same matter contents as the case of $T[J_{K_{1}}(n,q)]$ but with different mixed CS levels
\begin{itemize}[leftmargin=5mm]
\item UV (mixed) CS levels for $U(1)^2$ $-$ $U(1)^2$ : 
$
\begin{pmatrix}
\frac{1}{2}	&0	\\
0		&-\frac{1}{2}
\end{pmatrix}
$
\item UV mixed CS levels for $U(1)^2$ $-$ $U(1)_R$ : $\left(\frac{5}{2},\frac{3}{2}\right)$
\item UV mixed CS levels for  $U(1)^2$ $-$ $U(1)_v$ : $\left(\frac{1}{2}, -\frac{1}{2}\right)$
\item vortex line for $U(1)_v$ symmetry with the vortex charge $n$.
\end{itemize}

%%%%%%%%%%%%%%%%%%%%%%%%%%%%%%%%%%%%%%%%%%%%%%%%%%%%%%%%%%%%%%%%%%%%%%%%%%%%%%%%%%%%
%%%%%%%%%%%%%%%%%%%%%%%%%%%%%%%%%%%%%%%%%%%%%%%%%%%%%%%%%%%%%%%%%%%%%%%%%%%%%%%%%%%%
%%%%%%%%%%%%%%%%%%%%%%%%%%%%%%%%%%%%%%%%%%%%%%%%%%%%%%%%%%%%%%%%%%%%%%%%%%%%%%%%%%%%

\section{Remarks on quiver form and Neumann boundary condition}
\label{app:neumann}

We have mainly focused on the Dirichlet boundary conditions on vector multiplets and chiral multiplets.
In this section, we discuss an interpretation of the quiver form as a half-index of the 3d $\mathcal{N}=2$ theory with the Neumann boundary conditions for vector multiplets.
We take a quiver form of the type 
\begin{align}
\sum_{m_1, \ldots m_L \geq 0} q^{\frac{1}{2} \sum_{i,j} C_{ij} m_i m_j}  \prod_{j=1}^{L} \frac{w_j^{m_j}}{(q;q)_{m_j}}	\,	.
\label{ngendt}
\end{align}
The variable $w_j$ is given by, for example, $w_j = (-q^{\frac{1}{2}})^{p_j} a^{s_j} y^{h_j}$ where $p_j, s_j, h_j \in \mathbb{Z}$ for the generating function of HOMFLY polynomial.	\\

With the Neumann boundary conditions on vector multiplets, the half-index is given by the integral \eqref{nvint}.
We are interested in engineering the integral in such a way that the evaluation of the integral gives the quiver form \eqref{ngendt}.
For example, we can have the integral
\begin{align}
\begin{split}
&(q;q)_\infty^L \int_{\Gamma} \bigg( \prod_{j=1}^{L} \frac{d z_j}{ 2 \pi i z_j} \bigg)
\Bigg( \prod_{j=1}^L (qz_j;q)_\infty 
\theta(z_j;q)^{B_j+1}
\theta(z_j w_j ;q)^{-1} \Bigg)
\Bigg( \prod_{i < j} \theta(z_i z_j;q)^{-C_{ij}} \Bigg) 
\theta(1 ;q)^{-\sum_{j=1}^L \frac{B_j}{2}+1}
\end{split}	\label{ninteg0}
\end{align}
where $|q|<1$, $B_i := \big( \sum_{j=1}^{L} C_{ij} \big) -2C_{ii}$, $i=1, \ldots, L$, and $\theta(x;q)=((-q^{\frac{1}{2}})x;q)_\infty ((-q^{\frac{1}{2}})x^{-1};q)_\infty$.\footnote{We can choose another integral, for example,
\begin{align}
\begin{split}
(q;q)_\infty^L \int_{\Gamma} \bigg( \prod_{j=1}^{L} \frac{d z_j}{ 2 \pi i z_j} \bigg)
\Bigg( \prod_{j=1}^L (qz_j;q)_\infty \Bigg) \Bigg( \prod_{i < j} \bigg(\frac{\theta(1;q)}{\theta(z_i z_j;q)}\bigg)^{C_{ij}} \Bigg) 	 
\Bigg( \prod_{j=1}^L  \frac{\theta(w_j z_j;q)^{B_j+1}}{\theta(w_j^{B_j+2} z_j;q)} \frac{\theta(w_j^{B_j+2};q)}{\theta(w_j;q)^{B_j+1}}   \Bigg)	\,	.
\end{split}	\label{ninteg01}
\end{align}
After following the same procedure described below, this also gives \eqref{nresult} when $|q|>1$.
Therefore, the 3d $\mathcal{N}=2$ theory with boundary conditions extracted from \eqref{ninteg01} would be dual to the theory for \eqref{ninteg0} described in this section.
}
The 3d $\mathcal{N}=2$ theory whose half-index is \eqref{ninteg0} is given by
\begin{itemize}[leftmargin=5mm]
\item $U(1)^L$ vector multiplets with the $\mathcal{N}$ boundary condition
\item $L$ chiral multiplets $\Phi_i$, $i=1, \ldots, L$, of R-charge 0 and charge $-\delta_{ij}$ under $U(1)_j$, $j=1, \ldots, L$ with the D boundary condition 

\item For $B_i \geq 0 $

\hspace{1mm} $B_i+1$ 2d Fermi multiplets of R-charge $0$ and charge $\delta_{ij}$ under $U(1)_{j}$, $j=1, \ldots, L$

\vspace{0.5mm}

For $B_i \leq -2 $

\hspace{1mm} $-(B_i+1)$ 2d chiral multiplets of R-charge $1$ and charge $\delta_{ij}$ under $U(1)_{j}$, $j=1, \ldots, L$

\item $L$ 2d chiral multiplets, $i=1, \ldots, L$, of R-charge $1$ and charge $\delta_{ij}$ under $U(1)_j$ and $U(1)_{w_j}$, $j=1, \ldots, L$

\item For $C_{ij} \geq 1$

\hspace{1mm} $C_{ij}$ 2d chiral multiplets of R-charge $1$ and charge $(0, \ldots, \underset{i\text{-th}}{1} , \ldots, \underset{j\text{-th}}{1}, \ldots, 0)$ under $U(1)^L$

\vspace{0.5mm}

For $C_{ij} \leq -1$

\hspace{1mm} $-C_{ij}$ 2d Fermi multiplets of R-charge $0$ and charge $(0, \ldots, \underset{i\text{-th}}{1} , \ldots, \underset{j\text{-th}}{1}, \ldots, 0)$ under $U(1)^L$

\item For $\sum_{j=1}^L \frac{B_j}{2}+1 \geq 1 $,

\hspace{1mm} $\sum_{j=1}^L \Big( \frac{B_j}{2}+1 \Big)$ 2d chiral multiplets charged $(0,0,1)$ under $U(1)^L \times U(1)_{\mathbf{w}} \times U(1)_R$

For $\sum_{j=1}^L \frac{B_j}{2}+1 \leq -1 $,

\hspace{1mm} $-\sum_{j=1}^L  \Big(\frac{B_j}{2}+1 \Big)$ 2d Fermi multiplets charged $(0,0,0)$ under $U(1)^L \times U(1)_{\mathbf{w}} \times U(1)_R$

\item UV mixed CS levels for $U(1)^L$ $-$ $U(1)^L$ : $C_{ij}- \frac{1}{2} \delta_{ij}$
\item UV mixed CS levels for $U(1)^L$ $-$ $U(1)_R$ : $-\frac{1}{2}$
\item UV mixed CS levels for $U(1)_i$ $-$ $U(1)_{w_j}$, $i,j=1, \ldots, L$ : $\delta_{ij}$
\item UV CS level for $U(1)_R$ : $-L$

\end{itemize}
For the case of the generating function of the HOMFLY polynomial where the variable $w_j$ takes a form of $w_j = (-q^{\frac{1}{2}})^{p_j} a^{s_j} y^{h_j}$, the 2d chiral multiplets charged $l$ under $U(1)_{w_j}$ above are charged $(s_jl,h_jl)$ under $U(1)_a \times U(1)_y$ and the R-charge is added by $+p_jl$. 
The UV mixed CS levels for $U(1)_j$ -- $U(1)_R$, $U(1)_a$, and $U(1)_y$ become $p_j-\frac{1}{2}, s_j$, and $h_j$, respectively, $j=1, \ldots, L$.

When performing the integration, the convergent contour $\Gamma$ should be chosen, which corresponds to a vacuum of the 3d $\mathcal{N}=2$ theory under consideration.
The integrand in \eqref{ninteg0} is engineered to obtain \eqref{ngendt}, but for $|q|<1$, we don't have a good contour prescription that gives the quiver form \eqref{ngendt} analytically.
Instead, we may consider the half-index for $|q|>1$, which we often call the anti-half-index.
When $|q|>1$, the integral \eqref{ninteg0} is expressed as \cite{Beem-Dimofte-Pasquetti}
\begin{align}
\begin{split}
&(q^{-1};q^{-1})_\infty^{-L} \int_{\Gamma} \bigg( \prod_{j=1}^{L} \frac{d z_j}{ 2 \pi i z_j} \bigg)
\Bigg( \prod_{j=1}^L (z_j;q^{-1})_\infty 
\theta(z_j;q^{-1})^{B_j+1}
\theta(z_j w_j ;q^{-1})^{-1} \Bigg)^{-1}	\\
&\hspace{50mm} \times 
\Bigg( \prod_{i < j} \theta(z_i z_j;q^{-1})^{C_{ij}} \Bigg) 
\theta(1 ;q^{-1})^{\sum_{j=1}^L \frac{B_j}{2}+1}	\,	.
\end{split}	\label{ninteg}
\end{align}
We would like to choose a contour $\Gamma$ to pick poles at $z_j=q^{k}$, $k =0, 1, \ldots$, from $(z_j;q^{-1})_\infty^{-1}$, $j=1, \ldots, L$.
The asymptotics of the integrand of \eqref{ninteg} are given by
\begin{align}
\begin{array}{l l}
\exp \frac{1}{\hbar} \Big(\frac{1}{2} \sum_{j=1} (C_{jj} -1) (\log z_j)^2 + \sum_{j=1}^{L} \big(\log (-1) + \sum_{i \neq j} C_{ij} \log z_i + \log w_j \big) \log z_j \Big)	\,	,	&	\log |z_j| \rightarrow \infty	\vspace{1mm}\\
\exp \frac{1}{\hbar} \Big(\frac{1}{2}  \sum_{j=1} C_{jj} (\log z_j)^2 + \sum_{j=1}^{L} \big(\sum_{i \neq j} C_{ij} \log z_i + \log w_j \big) \log z_j \Big)	\,	,	&	\log |z_j| \rightarrow -\infty	\,	.
\end{array}
\end{align}
We can take $\hbar$ to be real, then $|q|>1$ means $\hbar >0$.
With an appropriate adjacency matrix $C_{ij}$ containing enough number of entries that are less than or equal to 1, we can choose a convergent contour $\Gamma$ that captures poles at $z=q^k$, $k=0, 1, \ldots$ and extends to $z_j \rightarrow \infty$.
However, in general there are cases that $C_{ij}$'s contain positive integers greater than 1.
Thus, for the cases with a large number of entries greater than 1, it may not be possible to have such a convergent contour.
Meanwhile, when considering the generating function of the HOMFLY polynomial with the totally symmetric representation $\mathcal{S}^r$, we can incorporate the framing $f \in \mathbb{Z}$ of a knot.
For the framing $f$, the factor $a^{fr} q^{\frac{1}{2}fr(r-1)}$ is multiplied to the summand of \eqref{genhomfly}, and the adjacency matrix $C_{ij}$ in the quiver form is shifted to \cite{KRSS2}
\begin{align}
C \rightarrow C+ f 
\begin{pmatrix} 
1	&1	&\cdots	&1	\\
1	&1	&\cdots	&1	\\
\vdots	&\vdots	&\ddots	&\vdots	\\
1	&1	&\cdots	&1	
\end{pmatrix}	\,	.
\end{align}
With some negative $f$, it would be possible to have suitable $C_{ij}$'s where we can choose a convergent contour.

The sum of the residues from the poles gives \eqref{ngendt} but with $|q|>1$,
\begin{align}
&\sum_{m_1, \ldots m_L \geq 0} (q^{-1})^{-\frac{1}{2} \sum_{i,j} (C_{ij}- \delta_{ij}) m_i m_j} (-q^{-\frac{1}{2}})^{\sum_{j} m_j}  \prod_{j=1}^{L} \frac{w_j^{m_j}}{(q^{-1};q^{-1})_{m_j}}	\\
&=\sum_{m_1, \ldots m_L \geq 0} q^{\frac{1}{2} \sum_{i,j} C_{ij} m_i m_j}  \prod_{j=1}^{L} \frac{w_j^{m_j}}{(q;q)_{m_j}}
\label{nresult}
\end{align}
with an overall factor $(q^{-1};q^{-1})_\infty^{-2L}$.

Given a proper expression \eqref{ngendt} for $|q|<1$, if \eqref{nresult} is the appropriate quantity for the region $|q|>1$, then the 3d $\mathcal{N}=2$ theory obtained above would be a theory whose anti-half-index is \eqref{nresult} and half-index is \eqref{ngendt} with appropriate choices of convergent contours for both.
For example, it is the case for the unknot whose generating function of HOMFLY polynomial is given by \eqref{homflygen01}.\footnote{More precisely, though we don't specify here a convergent contour for the case of the half-index \eqref{ninteg0} with $|q|<1$, considering that \eqref{homflygen01} can be expressed as a product of two $q$-Pochhammer symbols and the index on $S^2 \times_q S^1$ is obtained by a product of half-index and anti-half-index, we expect that there is an appropriate convergent contour that gives the half-index \eqref{ninteg0} with $|q|<1$ for the unknot. 
Then the 3d $\mathcal{N}=2$ theories for the unknot obtained in this section with the Neumann boundary condition and in section \ref{ssec:lk} with the Dirichlet boundary conditions for vector multiplets would be dual up to free chiral multiplets as their half-indices agree up to an overall factor $(q;q)_\infty^{2L}$.}
However, in general, there are some subtleties to consider.
The \eqref{nresult} with $|\tilde{q}|>1$ where we use the notation $\tilde{q}$ instead of $q$ for clarity is obtained formally by taking $q \rightarrow \tilde{q} = q^{-1}$ in \eqref{ngendt} with $|q|<1$.
The expression \eqref{ngendt} in general wouldn't be convergent for $|q|<1$ or $|q|>1$.
In addition, as mentioned in the footnote 12, it is known that there is an ambiguity to determine the appropriate series for $|q|>1$ by taking $q \rightarrow q^{-1}$ to the $q$-hypergeometric series with $|q|<1$ \cite{CCFGH}.
Therefore, even with $C_{ij}$'s that give convergent series both for  $|q|<1$ and $|q|>1$, it is unsure at the moment whether \eqref{nresult} with $|q|>1$ in general would be the appropriate quantity to consider as the anti-half-index from which the 3d $\mathcal{N}=2$ theory is obtained.	\\

The 3d $\mathcal{N}=2$ theories $T[Q_K]$ for the quiver $Q_K$ or $T[L_K]$ with the Neumann boundary conditions on vector multiplets corresponding to the generating function of the DT invariants or the HOMFLY polynomials have also been discussed in \cite{Ekholm-Kucharski-Longhi1}.
The information except $U(1)_R$ symmetry of the 3d $\mathcal{N}=2$ theory $T[L_K]$ was obtained from the asymptotic expansion in \cite{Ekholm-Kucharski-Longhi1}, and it agrees with the result in section \ref{ssec:lk} and in this section.
However, the information for $U(1)_R$ symmetry and also the boundary conditions were not specified in \cite{Ekholm-Kucharski-Longhi1}.
They can be read off from the integral expression for the half-index as discussed above, but there is an unusual relative minus sign for the fugacities $y_i$ in the $q$-Pochhammber symbol and the theta function in the integral expression eq(3.38) of \cite{Ekholm-Kucharski-Longhi1}.
Such relative minus sign doesn't arise naturally in the half-index of 3d $\mathcal{N}=2$ theories, but this might be accounted, for example, by introducing a fugacity, say $u$, for an additional $U(1)_u$ global symmetry that is added and then setting $u$ to a specialization $-1$ or to a $\mathbb{Z}_2$ torsion point \cite{Jockers:2021omw}.
Due to this relative minus sign, the factor $(-q^{-1}, q^{-1})_\infty$ arises after the evaluation of eq(3.38) in \cite{Ekholm-Kucharski-Longhi1}, which doesn't arise in the anti-half-index \eqref{nresult} of the 3d $\mathcal{N}=2$ theory with boundary conditions obtained in this section.
Thus, the quiver form \eqref{nresult} with overall factors containing $(-q^{-1}, q^{-1})_\infty$ arises from the anti-half-index at a special point of the fugacity $u$, \textit{i.e.} $u=-1$ for the theory extracted from eq(3.38) of \cite{Ekholm-Kucharski-Longhi1} with the $U(1)_u$ symmetry, while the quiver form \eqref{nresult} with the overall factor $(q^{-1};q^{-1})_\infty^{-2L}$ arises from the anti-half-index of the theory obtained in this section without introducing and tuning an additional parameter.\footnote{The theory obtained above could be dual to the theory extracted from eq(3.38) of \cite{Ekholm-Kucharski-Longhi1} with the $U(1)_u$ symmetry described above, if adding some number of matter multiplets, parts of which are charged under the $U(1)_u$ global symmetry introduced additionally, and mixed Chern-Simons coupling between the gauge symmetries and the $U(1)_u$ symmetry with appropriate identification of fugacities.
By taking a specialization of the fugacity $u$ to $-1$ in the anti-half-index of such theory with additional ingredients, the quiver form with overall factors containing $(-q^{-1}, q^{-1})_\infty$ are obtained.}	\\

As we see above, there are some subtleties in the interpretation of the quiver form as the half-index of 3d $\mathcal{N}=2$ theories with the Neumann boundary conditions on vector multiplets.
With Dirichlet boundary conditions, it seems that it is relatively more direct to obtain explicit 3d $\mathcal{N}=2$ theories with boundary conditions whose half-index is the quiver form under consideration.

%%%%%%%%%%%%%%%%%%%%%%%%%%%%%%%%%%%%%%%%%%%%%%%%%%%%%%%%%%%%%%%%%%%%%%%%%%%%%%%%%%%%
%%%%%%%%%%%%%%%%%%%%%%%%%%%%%%%%%%%%%%%%%%%%%%%%%%%%%%%%%%%%%%%%%%%%%%%%%%%%%%%%%%%%
%%%%%%%%%%%%%%%%%%%%%%%%%%%%%%%%%%%%%%%%%%%%%%%%%%%%%%%%%%%%%%%%%%%%%%%%%%%%%%%%%%%%

\section{3d $\mathcal{N}=2$ theories from the equivalent quivers}
\label{app:equivq}

It is known that more than one quiver gives the same DT generating function.
Such quivers are called the equivalent quivers \cite{KRSS2, Ekholm-Kucharski-Longhi2, JKLNS}.
Since their DT generating functions are the same, the corresponding 3d $\mathcal{N}=2$ theories are expected to be dual.

There are two kinds of quiver equivalences for symmetric quivers, one of which has a different size of the quiver and another of which preserves the size of the quiver.
Regarding the quiver equivalence that has a different size of the quiver, there are operations called unlinking, linking, and removing (or adding) redundant pairs \cite{Ekholm-Kucharski-Longhi2, EGGKPSS}.\footnote{A 3d $\mathcal{N}=2$ theory interpretation of unlinking and linking was discussed in \cite{Cheng:2023ocj}.}
\begin{itemize}[leftmargin=5mm]
\item unlinking

The unlinking of two nodes $a, b \in I$ in the quiver $Q=(I,A)$ takes a quiver $Q$ to another quiver $Q'=(I \cup n, A \cup \{\alpha_{ni}, \alpha_{in}\}_{i \in I})$ by adding a single extra node $n$ with arrows such that
\begin{align*}
C^{'}_{nn} &= C_{aa} + 2 C_{ab} + C_{bb}-1	\,	,	&&\hspace{-3mm}C^{'}_{ni} = C_{ai} + C_{bi} - \delta_{ai} - \delta_{bi}	\,	,	\\
C^{'}_{ab} &= C_{ab}-1	\,	,					&&\hspace{-2.5mm}C^{'}_{ij} =C_{ij}	\,	, \hspace{3mm} i,j \in I	\,	,
\end{align*}
where $\mathrm{x}'_n=q^{-1} \mathrm{x}_a \mathrm{x}_b$ and $\mathrm{x}'_i = \mathrm{x}_i$ for all $i \in I$.

\item linking

This is similar to the case of the unliking. 
The linking of two nodes $a,b \in I$ in the quiver $Q=(I,A)$ gives a new quiver $Q'=(I \cup n, A \cup \{\alpha_{ni}, \alpha_{in}\}_{i \in I})$ by adding a single extra node $n$ with arrows such that
\begin{align*}
C^{'}_{nn} &= C_{aa} + 2 C_{ab} + C_{bb}	\,	,		&&\hspace{-3mm}C^{'}_{ni} = C_{ai} + C_{bi}	\,	,	\\
C^{'}_{ab} &= C_{ab}+1	\,	,					&&\hspace{-2.5mm}C^{'}_{ij} =C_{ij}	\,	, \hspace{3mm} i,j \in I	\,	,
\end{align*}
where $\mathrm{x}'_n = \mathrm{x}_a \mathrm{x}_b$ and $\mathrm{x}'_i = \mathrm{x}_i$ for all $i \in I$.

\item removing redundant pairs

The pair $a$ and $b$ of nodes that satisfy $\mathrm{x}_a = q \mathrm{x}_b$, $C_{aa}=C_{ab}=C_{bb}-1$, and $C_{ai}=C_{bi}$ for all $i \in I \backslash \{a, b \}$ are called the redundant pair and can be removed from the quiver $Q=(I,A)$.

\end{itemize}

\vspace{3mm}

The quiver equivalence that preserves the size of the quiver has been discussed in \cite{JKLNS}.
\begin{itemize}[leftmargin=5mm]
\item quiver equivalence with the same size

Two quivers $Q$ and $Q'$ with $\mathrm{x}_i=\mathrm{x}^{'}_{i}$ are equivalent if they are related by the disjoint transposition of non-diagonal elements of $C$
\begin{align}
C_{ab} \leftrightarrow C_{cd}	\,	,	\qquad	C_{ba} \leftrightarrow C_{dc}
\end{align}
for pairwise different $a,b,c,d \in I$ such that
\begin{align}
\mathrm{x}_a \mathrm{x}_b = \mathrm{x}_c \mathrm{x}_d
\end{align}
and
\begin{align}
C_{ab} &= C_{cd} - 1	\,	,	\qquad	C_{ai}+C_{bi} = C_{ci} + C_{di} -\delta_{ci} -\delta_{di}	\quad \text{for any $i \in I$}	\\
&\hspace{20mm} \text{or}	\nonumber\\
C_{ab} &= C_{cd} +1	\,	,	\qquad	C_{ai}+C_{bi} = C_{ci} + C_{di} +\delta_{ai} +\delta_{bi}	\quad \text{for any $i \in I$}	
\end{align}
\end{itemize}

Equivalent quivers give the same quiver forms but with different expressions.
Therefore, 3d $\mathcal{N}=2$ theories whose half-indices are given by the equivalent quiver forms are expected to be dual to each other, and this applies to the 3d $\mathcal{N}=2$ theories obtained from the quiver form \eqref{gendt} or \eqref{gendtn}, for examples, which are discussed in section \ref{ssec:lk} and section \ref{sssec:afk}.

\end{appendices}

\bibliographystyle{JHEP}
\bibliography{ref}

\end{document}